\documentstyle[aps,prd]{revtex}
\input epsfig.sty 
\def\laq{\raise 0.4ex\hbox{$<$}\kern -0.8em\lower 0.62 ex\hbox{$\sim$}}
\def\gaq{\raise 0.4ex\hbox{$>$}\kern -0.7em\lower 0.62 ex\hbox{$\sim$}}

\begin{document}
\bibliographystyle{unsrt}

\title{Fully Anisotropic String Cosmologies, Maxwell Fields and Primordial 
Shear}

\author{Massimo Giovannini\footnote{Electronic address: 
m.giovannini@damtp.cam.ac.uk, giovan@cosmos2.phy.tufts.edu }}
\address{{\it Institute of Cosmology, Department of Physics and Astronomy,}}
\address{{\it Tufts University, Medford, Massachusetts 02155, USA}}

\maketitle

\begin{abstract}

We present a class of 
{\em exact} cosmological solutions of the low energy string 
effective action in the presence of a homogeneous magnetic fields. We 
discuss the physical properties of the obtained  (fully anisotropic)
 cosmologies
paying particular attention to their vacuum limit and to the possible
isotropization mechanisms. We argue that quadratic curvature corrections 
are able to isotropize fully anisotropic solutions
 whose scale factors describe accelerated expansion. 
Moreover, the degree of isotropization grows 
with the duration of the string phase. We follow the fate of the shear 
parameter in a decelerated phase  where, dilaton, magnetic fields and radiation
fluid are simultaneously present. In the absence of any magnetic field
 a {\em long string phase} 
immediately followed by radiation is able 
to erase large anisotropies. Conversely, if a 
{\em short string  phase} is followed
 by a long dilaton dominated phase the anisotropies can be present, 
in principle, also at later times. 
The presence of magnetic seeds after the end of the string phase
can induce further anisotropies which can be studied within 
the formalism reported in this paper.

\end{abstract}
\noindent

\renewcommand{\theequation}{1.\arabic{equation}}
\setcounter{equation}{0}
\section{Formulation of the Problem} 

Suppose that at some time $t_1$ the Universe becomes transparent to radiation 
and suppose that, at the same time, the four-dimensional background geometry 
(which we assume, for simplicity, spatially flat) has a very tiny 
amount of anisotropy in the scale factors associated with different spatial
 directions, namely
\begin{equation}
ds^2 = dt^2 - a^2(t)dx^2 - b^2(t) [ dy^2 + dz^2],
\label{1}
\end{equation}
where $b(t)$, as it will be clear in a moment, has to be only 
slightly different from $a(t)$. The electromagnetic radiation 
propagating along the $x$ and $y$ axes will have a  different
 temperature, namely  \cite{1,2} 
\begin{equation}
T_{x}(t) = T_1 \frac{a_1}{a} = T_1 e^{- \int H(t) dt},~~~
T_{y}(t) = T_1 \frac{b_1}{b} = T_1 e^{- \int F(t) dt},
\label{2}
\end{equation}
where $H(t)$ and $F(t)$ are the two Hubble factors 
associated, respectively with $a(t)$ and $b(t)$. 
The temperature anisotropy associated with the electromagnetic radiation 
propagating along two directions with different expansion rates can be 
roughly estimated, in the limit $H(t)-F(t)\ll 1$, as 
\begin{equation}
\frac{\Delta T}{T} \sim \int [H(t) - F(t)] dt= \frac{1}{2} 
\int \epsilon(t) d\log{t}
\end{equation}
where, in the second equality, we assumed that the deviations from the 
radiation dominated expansion can  be written as 
$F(t) \sim ( 1 - \epsilon(t))/2t$ with $|\epsilon(t)|\ll 1$. As we will 
see in the following $\epsilon$ can be connected with the shear 
parameter.

The dynamical origin of the primordial
anisotropy in the expansion \cite{2}
 can be connected with the existence of a primordial
magnetic field (not dynamically generated but postulated 
as an initial condition) or with some other sources 
of anisotropic pressure 
 and, therefore, the possible bounds on the temperature
 anisotropy can be translated into bounds on the time evolution of the 
anisotropic scale factors.

\begin{figure}
\centerline{\epsfxsize = 7 cm  \epsffile{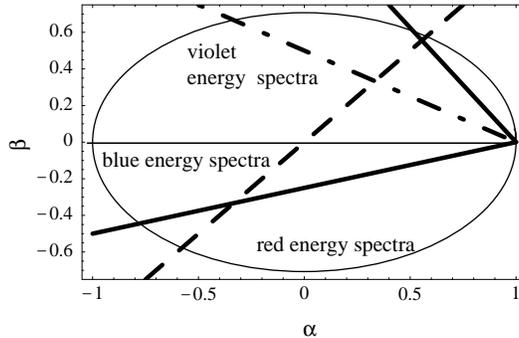}} 
\caption[a]{The axionic logarithmic energy spectra (in frequency)
are reported for different
dilaton-driven models with Kasner-like 
exponents $(\alpha,\beta)$ in the string frame. The models belonging to the 
third quadrant and localized on the arc of the ``vacuum'' ellipse 
$\alpha^2 + 2 \beta^2=1$ correspond 
to {\em anisotropic models with expanding and accelerated
 scale factors}.
If $\alpha$ and $\beta$ lie on the ellipse but 
 in the first quadrant we have solutions 
of the low energy beta functions which are both contracting 
(in the String frame). If $\alpha$ and $\beta$ on the ellipse but 
either in the second or in the fourth quadrant we have solutions of the 
low energy  beta functions where one of the two scale factors expands and the 
other contracts. Notice that the dashed (thick) line does correspond to the 
case of {\em fully isotropic} solutions (i.e. $\alpha = \beta = -1/\sqrt{3}$)
 whose intersection with the vacuum ellipse lies in the region of red spectra.
{\em Above} the dot-dashed (tick)
 line the dilaton {\em decreases} for $t<0$ whereas 
{\em below} the dot-dashed line the dilaton {\em increases}. We would be 
tempted to speculate that to have an increasing dilaton is  a {\em sufficient}
condition in order to have a pre-big-bang dynamics. This is {\em not} the case.
Indeed, an {\em increasing} dilaton is  also compatible with $\alpha$ and 
$\beta$ in the second quadrant where the scale factors are {\em not both
expanding}. Therefore, an increasing dilaton is not a sufficient condition
for anisotropic pre-big-bang dynamics but {\em only a necessary} condition. }
\label{f1}
\end{figure}

In  
the context of string cosmology  \cite{5} there are no reasons 
why the Universe should not be  mildly anisotropic
as it was recently pointed out \cite{6}.
Indeed, the Universe {\em can be} anisotropic as a result of the Kasner-like
nature of the pre-big-bang solutions. Moreover, we could argue, that 
the Universe {\em has to be only mildly anisotropic}. By mildly anisotropic
we mean that the scale factors should be {\em both expanding} 
(in the String frame). 
The low energy beta functions can be  solved for a metric 
(\ref{1}) with the result \cite{6}
\begin{equation}
 a(t) = \biggl[-\frac{t}{t_1}\biggr]^{\alpha},~~~ b(t) = 
\biggl[-\frac{t}{t_1}\biggr]^{\beta},~~~\phi(t) = ( \alpha + 2 \beta
-1)\log{\biggl[-\frac{t}{t_{1}}\biggr]},
\label{solt}
\end{equation}
where $\alpha^2 + 2 \beta^2 =1$. If we want $\dot{a}>0$ and $\dot{b}>0$
(expanding solutions) and $\ddot{a}>0$, $\ddot{b}>0$ (inflationary solutions)
the exponents $\alpha$ and $\beta$ have to lie in the third quadrant 
along the arc of the ellipse reported in Fig. \ref{f1}.
The requirement of having, in the same class of 
fully anisotropic models, axion spectra decreasing at large distances (i.e. 
blue frequency spectra) will select a slice exactly in the same region
of parameter space \cite{6}. Therefore, general considerations seem to 
point towards {\em mildly anisotropic pre-big-bang models}.
Since for phenomenological considerations \cite{7}
 we would like to have blue (or flat)
logarithmic energy spectra it seems reasonable to analyze 
fully anisotropic string cosmological models. 
Recently it was  pointed out that the collapse of a stochastic collection of
dilatonic waves (in the Einstein frame) 
does also lead to an anisotropic pre-big-bang phase \cite{8}.

In the framework of an anisotropic pre-big-bang phase it 
is  plausible, from a theoretical point of view, to investigate the 
role played by a homogeneous magnetic field which could represent 
a further source of anisotropy. Thus,
the first technical point we want to investigate is the possible 
generalization of the pre-big-bang solutions to the case of a homogeneous 
magnetic field. This generalization is not so straightforward since, 
in the low energy limit, the dilaton field is directly coupled to the kinetic 
term of the Maxwell field
\begin{equation}
S= - \frac{1}{2 \lambda_{s}^2} \int d^4 x \sqrt{- g} e^{-\phi} \biggl[ R + 
g^{\alpha\beta} \partial_{\alpha}\phi \partial_{\beta} \phi - 
\frac{1}{12} H_{\mu\nu\alpha} H^{\mu\nu\alpha} 
+ \frac{1}{4} F_{\alpha\beta}F^{\alpha\beta} +...\biggl]
\label{action}
\end{equation}
where $F_{\alpha\beta}= \nabla_{[\alpha} A_{\beta]}$ is the Maxwell 
field strength and $\nabla_{\mu}$ is the covariant derivative with respect 
to the String frame  metric $g_{\mu\nu}$. Notice that $H_{\mu\nu\alpha}$ is the antisymmetric field strength.
In Eq. (\ref{action}) 
the ellipses stand for a possible (non-perturbative) dilaton potential 
and for the string tension corrections parameterized by 
$\alpha' = \lambda_{s}^2$ (notice that $\lambda_{s}$ is the string scale). 
In Eq. (\ref{action}) $F_{\mu\nu}$ can be thought as the Maxwell field 
associated with a $U(1)$ subgroup of $E_{8}\times E_{8}$. One might think 
that by going to the Einstein frame the dilaton can be decoupled from 
the kinetic term of the Maxwell fields. This is not the case as we discuss, in 
greater detail, in Appendix A.

\begin{figure}
\begin{center}
\begin{tabular}{|c|c|}
      \hline
      \hbox{\epsfxsize = 6.5 cm  \epsffile{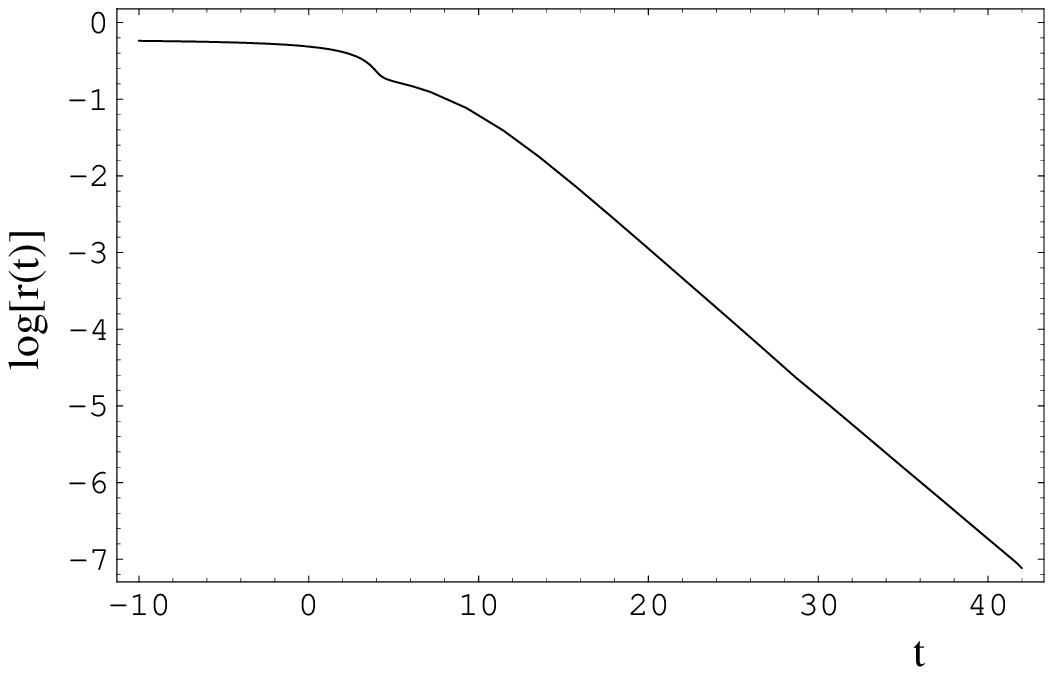}} &
      \hbox{\epsfxsize = 6.5 cm  \epsffile{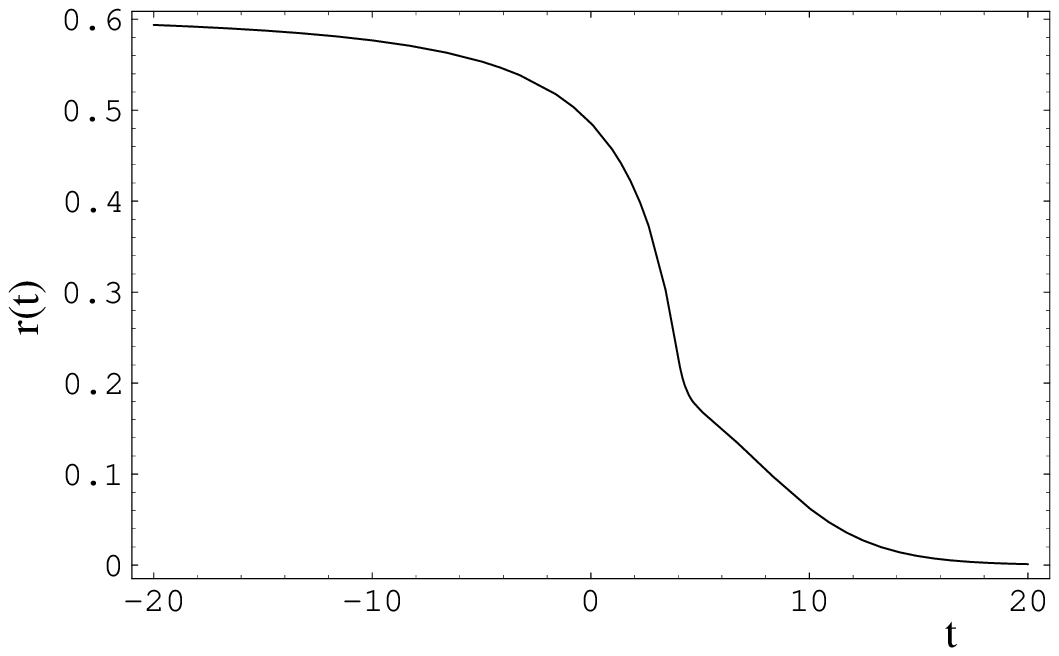}} \\
      \hline
\end{tabular}
\end{center}
\caption[a]{We plot the evolution of the degree of anisotropy (left) and of
the shear parameter itself (right) in the case 
of a mildly anisotropic solution with $\alpha = - 7/9$ and 
$\beta= - 4/9$ corresponding
 to the case of flat axionic spectra \cite{6} 
(i.e. the intersection of the thick full line with the 
vacuum ellipse in the third quadrant of Fig. \ref{f1}). As  we 
can see for $t\rightarrow - \infty$, $r(t) \rightarrow 
[-7/9 + 4/9]/[-7/9 - 4/9] = 3/11$. For 
$t\rightarrow + \infty$ the shear parameter and the degree 
of anisotropy gets reduced as the result of the addition of 
of higher order string tension corrections (see also Section IV). 
We notice that the degree of isotropization 
crucially depends upon the {\em duration} of the 
string phase (left).  In a {\em short} string phase $I(t)$ 
can be of the order of $-3$, $-4$. If 
the string phase is very long the degree of anisotropy can be 
much smaller than $-7$. Notice that, by only looking at 
the right picture we might guess (in a wrong way) that the 
shear parameter tends (for $t\rightarrow +\infty$) to a constant
(small) value. The crucial point we want to stress is that 
this is not the case as we can understand by looking  at $I(t)$
which  is {\em always a decreasing function of the cosmic time}.}
\label{f2}
\end{figure}

In addressing the possible occurrence of an anisotropic phase in the life
of the Universe there is an immediate concern. We want to make sure that 
the amount of anisotropy encoded in the initial conditions will be eventually
washed out by the subsequent evolution. It can be shown, in this 
respect,  that an interesting role can be played by the string tension 
corrections \cite{6}. In fact, by adding the first $\alpha'$ correction to 
the tree-level action reported in Eq. (\ref{action}) two interesting 
things happens. On one hand the curvature invariants get regularized and, on 
the other hand, an anisotropic solution with expanding scale factors gets 
isotropized. This statement can be made more precise by looking, 
simultaneously at Fig. \ref{f1} and Fig. \ref{f2}. In Fig. \ref{f2} we plot 
the shear parameter $r(t) = 3[ H(t) - F(t)]/ [ H(t) + 2 F(t)]$ for the
particular vacuum solution leading to flat axionic spectra in four 
anisotropic dimensions, namely the case $\alpha= -7/9$, $\beta = - 4/9$. 
This case does correspond in Fig. \ref{f1} to the intersection between the 
full line and the vacuum ellipse in the third quadrant. We can clearly see 
that for $t\rightarrow - \infty$ the solution is anisotropic since
$r(t)$ is of order one.  The crucial 
question for the purpose of the present paper 
 is by how much the quadratic corrections 
are able to reduce the degree of anisotropy. 

For our future convenience we {\em define} the degree 
of isotropization as the logarithm (in then basis) of the 
absolute value of the shear parameter:
\begin{equation}
I(t) = \log{|r(t)|}. ~~~
\end{equation}
The reason for the absolute value in the last equations 
is that $r(t)$ can change sign depending upon the relative magnitude 
of $H$ and $F$.
In Fig. \ref{f2} we plot the degree of isotropization (at the left)
and the shear parameter itself (at the right).
We see that the degree of isotropization is a function {\em of the duration 
of the string phase}. Thus if we have a very long string phase 
we can have $I(t) \ll -7$. On the other hand if the duration of the string
phase is not too long, then one can naturally have larger values for $r(t)$.
Suppose now that the string phase stops at some time $t$ 
and it is replaced by a radiation 
dominated phase. Then by the continuity of the scale factors at the 
transition time  the anisotropy in the expansion is translated in a 
temperature ansisotropy according to Eq. (\ref{1}). 
Therefore, it could happen, depending 
upon the duration of the string phase, that either $I(t) \gaq 10^{-5}$
or $I(t) \laq 10^{-5}$ right after the end of the string 
phase (under the assumption that radiation sets in instantaneously).
 Many questions can arise at this point. Will
the radiation evolution completely erase the primordial anisotropy? 
If this is 
not the case, should we exclude, on a phenomenological ground, too short
string phases since they might potentially conflict with the observed 
value of the large scale anisotropy? Which is the 
role of a primordial magnetic field in the picture? We want to recall, 
in fact,  that even if we would not include a magnetic field in the 
pre-big-bang phase such a field will be anyway dynamically generated by 
parametric amplification of the electromagnetic (vacuum) fluctuations 
\cite{9} and 
it will evolve, in the radiation dominated (post-big-bang) phase
 affecting the evolution of 
$r(t)$ \cite{1}. Certainly one can argue that the magnetic 
field  generated  from the electromagnetic vacuum fluctuations 
is stochastic in nature as all the other fields 
(gravitons, scalar fields...)
produced as the result of parametric amplification 
of quantum mechanical initial conditions 
\cite{9b,9c}. Therefore, we could conjecture that the
obtained magnetic field does not have 
any preferred direction. We want to point 
out that this conclusion has been reached by looking 
at the magnetic field generation in  a completely
homogeneous and isotropic dilaton-driven phase.
If the background {\em is not completely isotropic}
during the dilaton driven evolution, then, we can
argue, in analogy with the graviton production
in anisotropic backgrounds of Bianchi type I,
that the produced background has different 
properties in different spatial directions \cite{9d}.

Finally, the considerations and the various questions 
 we just stated hold in the case of 
transition from the string phase to the radiation dominated phase. However, 
it could happen that before reaching the radiation phase the background 
evolves toward a state of decreasing dilaton and then, we should treat  
again the same issues in the case of a background with decreasing 
dilaton coupling. These are the problems  we want to investigate in the 
second part of the present paper.

Our plan is the following. In Section II we will discuss the magnetic 
solutions of the low energy beta functions whereas in Section III we will
discuss their physical interpretation. In Section IV we will 
investigate the role of the $\alpha'$ corrections and we will try to follow
the evolution of the degree of anisotropy all along the string phase. 
In Section V we will study the  fate of the anisotropy in a decelerated phase 
and we will address the issue of possible bounds on the duration of the 
string phase from too large temperature anisotropies. 
Section VI contains our concluding remarks.
Some useful technical results concerning the description of anisotropies
and of magnetic fields in the String and in the Einstein frames 
are collected in Appendix A.

\renewcommand{\theequation}{2.\arabic{equation}}
\setcounter{equation}{0}
\section{Magnetic Solutions}

Consider a spatially flat background configuration with vanishing 
antysimmetric field strangth 
($H_{\mu\nu\alpha} =0$) and vanishing dilaton potential. 
The dilaton depends only on time and the metric will be taken fully 
anisotropic since we want to study possible solutions with a homogeneous 
magnetic field which is expected to break the isotropy of the background:
\begin{equation} 
ds^2 = g_{\mu\nu} dx^{\mu} dx^{\nu} = dt^2 - a^2(t) dx^2 - b^2(t) dy^2 
- c^2(t) dz^2 .
\label{metric}
\end{equation}
This choice of the metric corresponds to a synchronous coordinate system in 
which the time parameter coincides with the usual cosmic time 
(i.e. $g_{00} =1$ and $g_{0i} =0$).

By varying the effective action with respect to $\phi$, $g_{\mu\nu}$ and the 
vector potential $A_{\mu}$  we get, respectively 
\begin{eqnarray}
&&R - g^{\alpha\beta} \partial_{\alpha}\phi \partial_{\beta}\phi + 
2 g^{\alpha\beta} \nabla_{\alpha} \nabla_{\beta} \phi = 
- \frac{ 1}{4} F_{\alpha\beta} F^{\alpha\beta}, 
\label{phi}\\
&& R_{\mu}^{\nu} - \frac{1}{2} \delta_{\mu}^{\nu} R = \frac{1}{2} 
\biggl[ \frac{1}{4} \delta_{\mu}^{\nu} F_{\alpha\beta} F^{\alpha\beta} 
- F_{\mu\beta} F^{\nu\beta}  \biggr]
- \frac{1}{2} \delta_{\mu}^{\nu}  g^{\alpha\beta}\partial_{\alpha}\phi 
\partial_{\beta} \phi - \nabla_{\mu} \nabla^{\nu} \phi + \delta_{\mu}^{\nu} 
\Box\phi,
\label{g}\\
&& \nabla_{\mu}\biggl[ e^{-\phi} F^{\mu\nu}\biggr] =0.
\label{b}
\end{eqnarray}
where $\nabla_{\mu}$ denotes covariant differentiation with respect 
to the metric 
of Eq. (\ref{metric}). Inserting Eq. (\ref{phi}) into Eq. (\ref{g}) we 
get that Eq. (\ref{g}) 
can  be expressed as 
\begin{equation}
R_{\mu}^{\nu} + \nabla_{\mu}\nabla^{\nu} \phi 
+ \frac{1}{2} F_{\mu\alpha}F^{\nu\alpha} =0.
\label{g2}
\end{equation}
Consider now a homogeneous magnetic field directed along the $x$ axis. 
The generalized Maxwell equations (\ref{b}) 
and the associated Bianchi identities 
can then be trivially solved by  the field strength $F_{yz}= -F_{zy}$.
The resulting system of equations (\ref{phi}), (\ref{g2}) can than be 
written, in the metric of Eq. (\ref{metric}),  as
\begin{eqnarray}
&&2 \ddot\phi + 2 ( H + F + G) \dot\phi - \dot{\phi}^2 -
 2 \biggl[ H F + F G + F G + H^2 + F^2 + G^2 + \dot{H} + \dot{F}
 +\dot{G}\biggr]+ \frac{B^2}{2 b^2 c^2}= 0,
\label{phi1}\\
&&\ddot{\phi} = H^2 + F^2 + G^2 + \dot{H} + \dot{F} + \dot{G},
\label{00}\\
&&H\dot{\phi} = H F + H G +  H^2 + \dot{H},
\label{xx}\\
&&F\dot{\phi} = H F + F G + F^2 + \dot{F} - \frac{B^2}{2 b^2c^2},
\label{yy}\\
&&G \dot{\phi} = HF + F G + G^2 +\dot{G} - \frac{B^2}{2 b^2 c^2}, 
\label{zz}
\end{eqnarray}
where Eq. (\ref{phi1}) is the dialton equation and Eqs. 
(\ref{00})--(\ref{zz}) do correspond, respectively to the 
$(00)$ and $(ii)$  components of Eq. (\ref{g2}).
Notice that $B$ is the magnetic field intensity. The inclusion of a
 time-dependent $B$ is in principle possible but in order to 
be consistent with the Bianchi indentities we would need to include 
also an electric field (not necessarily homogeneous). Notice that 
in the limit $B\rightarrow 0$ we obtain the well known form of 
the low energy beta functions in  a fully anisotropic metric.

In order to give a general solution of the previous system it is convenient 
to define a generalized ``conformal'' time namely
\begin{equation}
e^{-\overline{\phi}} d \eta = dt,~~~~\overline{\phi} = \phi - \log{\sqrt{-g}},
\label{time}
\end{equation}
where $ \overline{\phi}$ is the ``shifted'' dilaton which is invariant 
for duality related vacuum solutions.
In terms of $\eta$ Eqs. (\ref{phi})--(\ref{zz}) become, respectively, 
\begin{eqnarray}
&&2 \phi'' + {\phi'}^2 - 2 \Sigma'
- 2 \phi' \Sigma  +   \Lambda = - \frac{B^2}{2} a^2 e^{- 2 \phi}, 
\label{phi2}
\\
&&\phi'' + {\phi'}^2 - 2 \Sigma \phi'  + \Lambda  - \Sigma' =0,
\label{002}
\\
&&{\cal H}' =0,~~~{\cal F}' = \frac{B^2}{2} a^2 e^{- 2\phi} 
,~~~{\cal G}' = \frac{B^2}{2} a^2 e^{- 2\phi}, 
\label{ii2}
\end{eqnarray}
where 
\begin{equation}
\Sigma = ( {\cal H} + {\cal F} + {\cal G}),~~~\Lambda = 2 ( {\cal H} {\cal F}+ 
{\cal H} {\cal G} + {\cal F}{\cal G}),
\end{equation}
with the obvious notation that $(\log{a})' = {\cal H}$, 
$(\log{b})' = {\cal F}$, $(\log{c})' = {\cal G}$  
(the prime denotes differentiation with respect to $\eta$).
By subtracting Eq. (\ref{002}) from Eq. (\ref{phi2}) we obtain 
\begin{equation}
\phi'' - \Sigma' =- \frac{B^2}{2} a^2 e^{- 2 \phi}. 
\label{diff}
\end{equation}
Using the remaining equations in Eq. (\ref{diff}) in order to eliminate 
$\Sigma'$ we get a general form of the solution which reads 
\begin{equation}
a(\eta) = a_{0} e^{{\cal H}_{0} \eta},~~~
b(\eta) = b_{0} e^{{\cal F}_{0} \eta}e^{\phi},~~~
c(\eta) = a_{0} e^{{\cal G}_{0} \eta}e^{\phi},~~~
\label{soltrial}
\end{equation}
where $\phi$ satisfies 
\begin{equation}
\phi'' + {\phi'}^2 - 2 {\cal H}_0 \phi'  - \Lambda_0 =0,
~~~\Lambda_0 = 2( {\cal H}_0 {\cal F}_0+
{\cal F}_0 {\cal G}_0 + {\cal F}_0 {\cal G}_0).
\label{eqtrial}
\end{equation}
By solving this last equation and by inserting the (trial)  
solution (\ref{soltrial}) into Eqs. (\ref{phi2})--(\ref{zz}) we will get 
a set of consistency relations among the different integration constants 
which will determine a general form of the solutions in terms of the
 (undetermined) set of initial conditions. 

Let us notice  that by  repeatedly combining the
 equations of motion we obtain a very useful relation, namely 
$2 \phi'' = (B a)^2 \exp{[- 2 \phi]}$. Since the left hand side of this
last equation is positive definite, we have also to demand, for physical
consistency that, from Eq. (\ref{eqtrial}),  $ -{\phi'}^2 
+ 2{\cal H}_0 \phi' + \Lambda_0 >0$, which also implies
\begin{equation}
{\cal H}_0 + \sqrt{{\cal H}_0^2 + \Lambda_0} < \phi' < {\cal H}_0 - 
\sqrt{{\cal H}_0^2 + \Lambda_0}.
\end{equation}
Notice that this is a requirement to be satisfied in the presence 
of a constant {\em magnetic} 
field and it does change in the presence of a constant 
{\em electric} field (indeed for a constant electric field directed along 
the $x$ direction and in the absence of any associated magnetic field 
we would have that, in the same conformal time parameterization $\phi''<0$).

The wanted solution is 
\begin{eqnarray}
&&a(\eta)= a_0 e^{{\cal H}_0 \eta},~~~b(\eta) = b_0 e^{\phi_0} 
e^{({\cal F}_0 + {\cal H}_0 + \Delta_0)\eta}\biggl[e^{-2 \Delta_0 \eta}
 + e^{\phi_1} \biggr],~~~c(\eta) = c_0 e^{\phi_0} 
e^{({\cal G}_0 + {\cal H}_0 + \Delta_0)\eta}\biggl[e^{-2 \Delta_0 \eta}
 + e^{\phi_1} \biggr],
\\
&& \phi(\eta) = \phi_0 + ({\cal H}_0 + \Delta_0)\eta + 
\log{\biggl[ e^{\phi_1} + e^ {- 2 \Delta_0 \eta}\biggr]},
\label{s1}
\end{eqnarray}
where, by consistency with all the other equations we have 
\begin{equation}
\Delta_{0} \equiv\sqrt{{\cal H}_0^2 + \Lambda_0}= \frac{a_0}{2 \sqrt{2}}
e^{- (\phi_0 + \phi_1)} B.
\label{c1}
\end{equation}
A particular solution of the system is fixed by fixing the value of 
the magnetic field in string units.
Before discussing the physical properties of the solution (\ref{s1})
 we want to see 
in which way the vacuum solutions appear in the time parameterization 
defined by Eq. (\ref{time}). The vacuum solutions are quite straightforward 
using $\eta$ and they correspond to constant Hubble factors and linear
 dilaton, namely 
\begin{equation}
{\cal H} = {\cal H}_0,~~~{\cal F} = {\cal F}_{0},~~~{\cal G}= {\cal G}_0,~~~
\phi(\eta) = \phi_0 + \phi_{2} \eta, 
\label{s2}
\end{equation}
 with the condition,
\begin{equation}
\phi_2 = {\cal H}_0 + {\cal F}_0 + {\cal G}_0 \pm \sqrt{ {\cal H}_0^2 
+ {\cal F}_0^2 + {\cal G}_0^2}.
\label{c2}
\end{equation}
It is straightforward to see that these solutions are indeed  Kasner-like 
by working out the cosmic time picture from Eq. (\ref{time}). Indeed, with 
little effort  we  can see that
\begin{equation}
a(t)= a_0 \biggl( \frac{t}{t_0}\biggr)^{\alpha_1},~~~
b(t)= b_0 \biggl( \frac{t}{t_0}\biggr)^{\alpha_2},~~~
c(t)= c_0 \biggl( \frac{t}{t_0}\biggr)^{\alpha_3},
\end{equation}
where
\begin{equation}
\alpha_1= \frac{ {\cal H}_0}{{\cal H}_0 + {\cal F}_0 + {\cal G}_0  - \phi_2}
,~~~\alpha_2=
 \frac{ {\cal F}_0}{{\cal H}_0 + {\cal F}_0 + {\cal G}_0  - \phi_2}
,~~~\alpha_3 =\frac{ {\cal G}_0}{{\cal H}_0 + {\cal F}_0 
+ {\cal G}_0  - \phi_2}
\end{equation}
Using Eq. (\ref{c2}) we can obtain that the three (cosmic time) 
exponents satisfy the Kasner-like condition, namely $\alpha_1^2 + \alpha_2^2
+ \alpha_3^2 =1$ and the $\pm$ ambiguity of Eq. (\ref{c2}) simply refers 
to the two (duality related) branches.

Before concluding the present Section we want to mention that the equations 
of motion can also  be integrated in the case where the coupling of 
the dilaton to the Maxwell fields is slightly different \cite{10} from the 
one examined in this Section, namely in the case where the action  
is written as 
\begin{equation}
S= - \frac{1}{2 \lambda_{s}^2} \int d^4 x \sqrt{- g} e^{-\phi} \biggl[ R + 
g^{\alpha\beta} \partial_{\alpha}\phi \partial_{\beta} \phi  
+ \frac{1}{4}e^{- q \phi} F_{\alpha\beta}F^{\alpha\beta} \biggl]
\label{action2}
\end{equation}
with $q\neq 0$. 

In this case the equations of motion in the generalized conformal time $\eta$
can be written as
\begin{eqnarray}
&& 2 \phi'' + {\phi'}^2 - 2 \Sigma\phi' +  \Lambda - 2 \Sigma' = 
- \frac{ q + 1 }{2} B^2 a^2 e^{- ( q + 2) \phi}
\\
&&\phi'' + {\phi'}^2 - 2 \Sigma \phi' - \Sigma' + \lambda = - \frac{ q}{4} 
B^2 a^2 e^{- (q + 2)\phi}
\\
&& {\cal H}' = \frac{ q }{4} B^2 a^2 e^{ - ( q + 2)\phi} ,~~~
{\cal F}' = \frac{q + 2}{4} B^2 a^2 e^{- (q + 2 )\phi},~~~
{\cal G}' = \frac{q + 2}{4} B^2 a^2 e^{- (q + 2 )\phi}.
\end{eqnarray}
Notice that in this case the system changes qualitatively but it can still 
be integrated. In fact, by following the procedure outlined in the previous 
paragraphs we obtain a decoupled equation for $\phi'$ 
\begin{equation}
- \frac{\phi''}{q + 1} - \frac{ q^2 + 2 q + 2}{ 2 (q + 1)^2} {\phi'}^2
+ \frac{2 {\cal H}_0}{ q + 1} \phi' + \Lambda_0=0
\end{equation}
This equation can be easily integrated providing the solutions 
to the whole system in the case $q \neq 0$. 

\renewcommand{\theequation}{3.\arabic{equation}}
\setcounter{equation}{0}
\section{Analysis of the Magnetic Solutions}

\begin{figure}
\begin{center}
\begin{tabular}{|c|c|}
      \hline
      \hbox{\epsfxsize = 6.5 cm  \epsffile{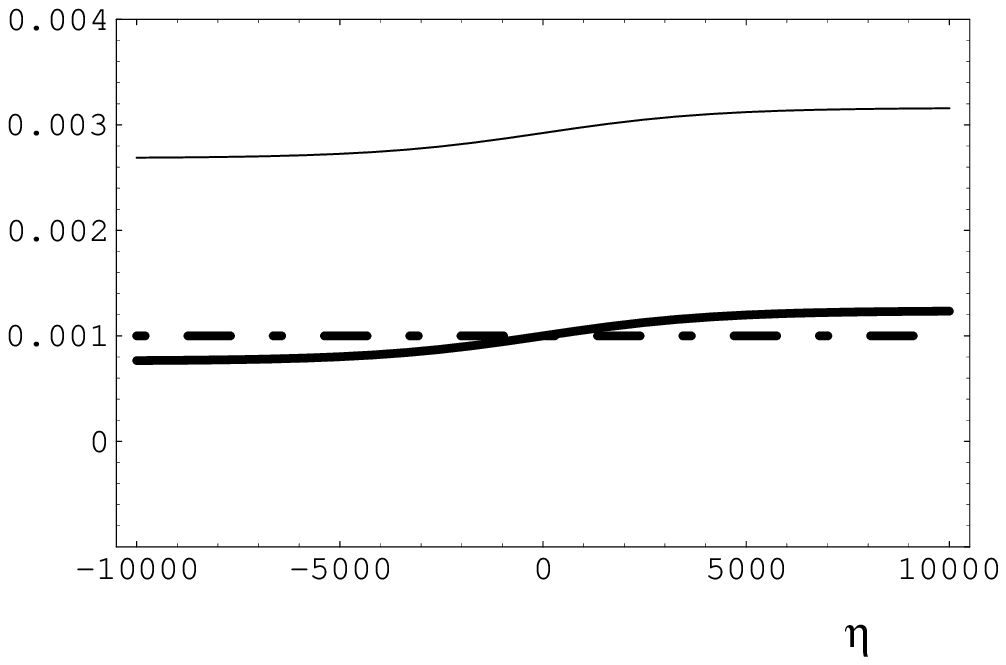}} &
      \hbox{\epsfxsize = 6.5 cm  \epsffile{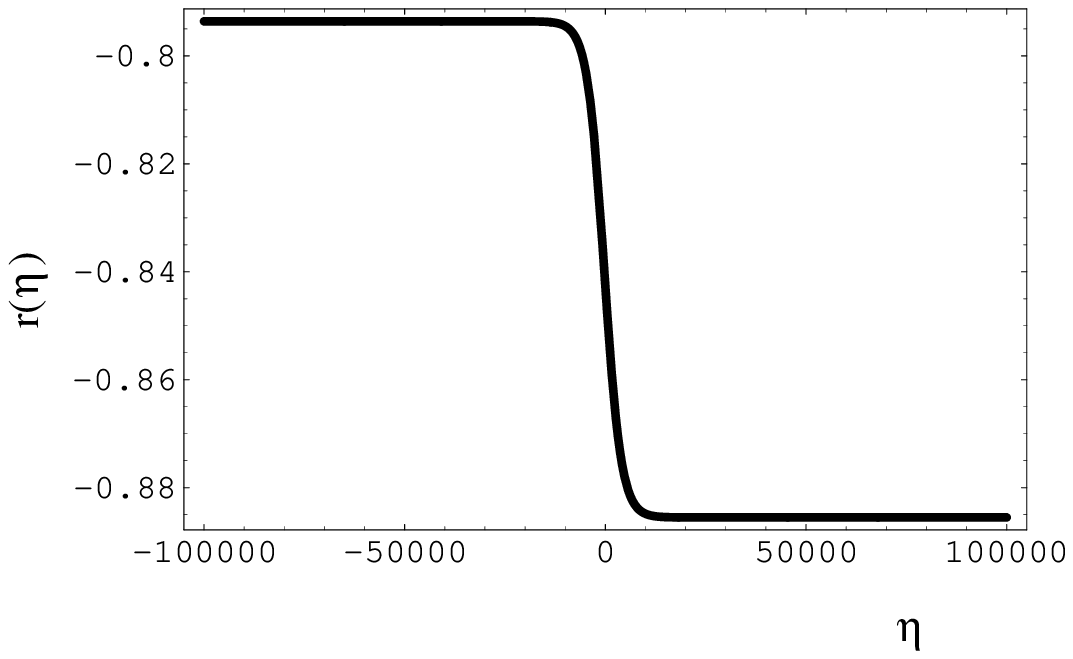}} \\
      \hline
\end{tabular}
\end{center}
\caption[a]{ We plot the magnetic solutions given in Eqs. (\ref{s1})
 in the case where 
${\cal H}_0=0.001$, $B=01$, $\phi_0=-5$. In the {\em left} plot we 
illustrate the behavior of ${\cal H}$
({\em dot-dashed line}), ${\cal F}$ ({\em thin line}) and $\phi'$ 
({\em full thick line}).
 At the right we 
illustrate the behavior of the shear parameter. We notice that,
 in spite of the appearance,
the shear parameter does not decrease, asymptotically but it tends to
 a constant 
value  of order one, in sharp contrast with what happens 
if we include quadratic corrections to the tree-level
 action (see next Section).
We see that ${\cal H}$, ${\cal F}$ and $\phi'$
are  {\em almost} constant if the magnetic field is {\em small }
 in String units. }
\label{f3a}
\end{figure}

\begin{figure}
\begin{center}
\begin{tabular}{|c|c|}
      \hline
      \hbox{\epsfxsize = 6.5 cm  \epsffile{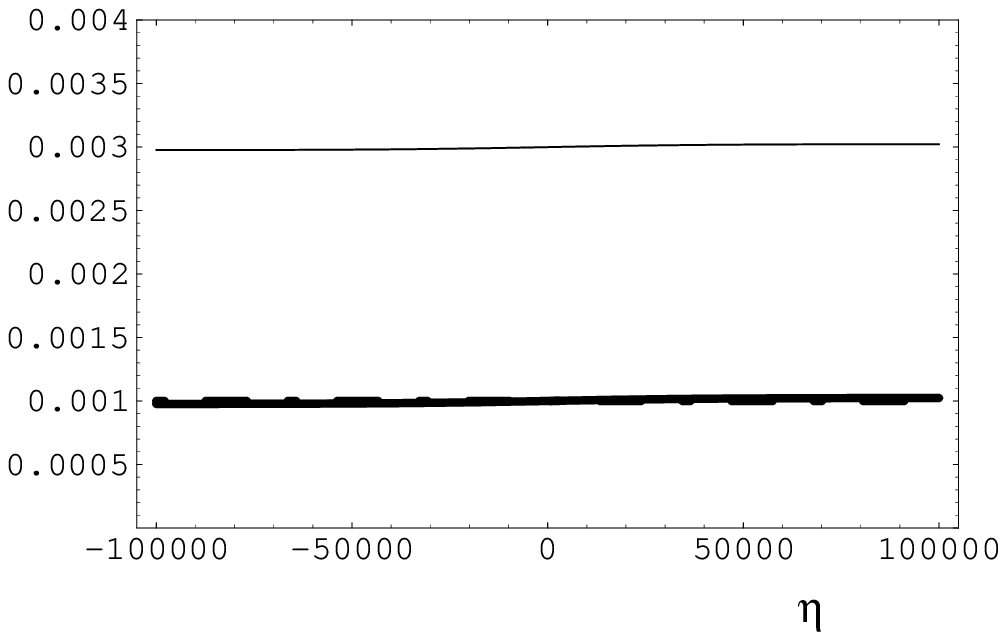}} &
      \hbox{\epsfxsize = 6.5 cm  \epsffile{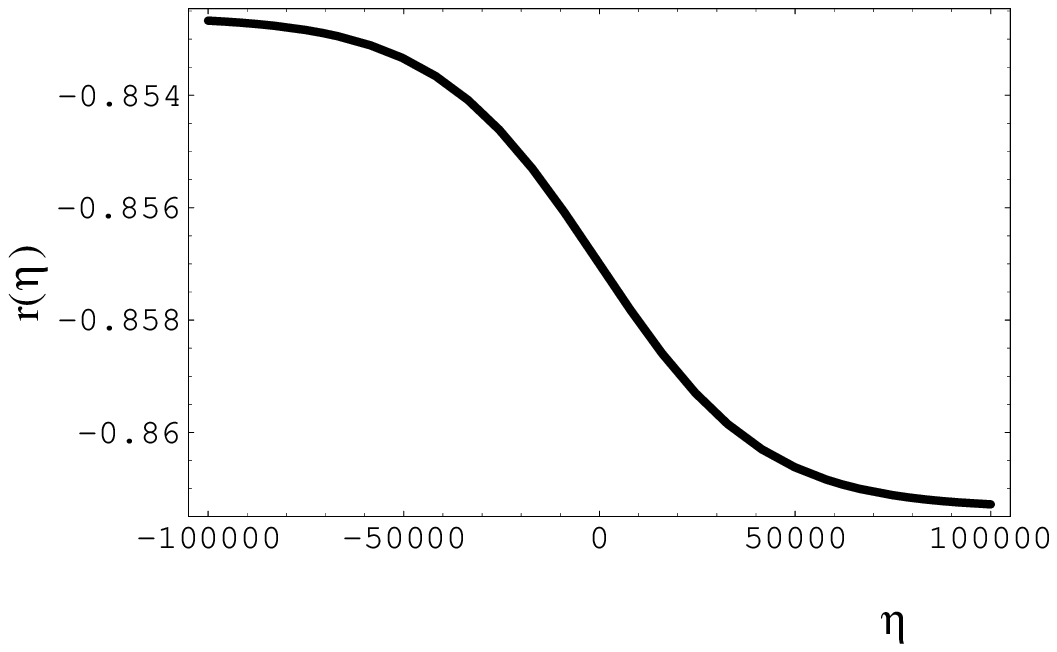}} \\
      \hline
\end{tabular}
\end{center}
\caption[a]{We plot the same solution given in
 Fig. \ref{f3a} but with a different 
value of the magnetic field which is now $B=0.001$. As in the previous plot at 
the {\em left} we report the evolution of ${\cal H}$ 
({\em dot-dashed line}), ${\cal F}$
 ({\em thin line})and $\phi'$ 
( {\em full thick line}). Notice that in this case the growth 
of $\phi'$ is so minute that the dot-dashed line and the full thick 
lines are almost indistinguishable. This is due to the fact that
$\phi'$ continuously interpolates between ${\cal H}_0 - \Delta_0$ and 
${\cal H}_0 + \Delta_0$ and $\Delta_0 \sim 0.001 e^{-5}$. At the {\em right}
we report the evolution of the shear parameter for the same solution.}
\label{f3b}
\end{figure}

In order to analyze the previous solution let us assume, for sake of 
simplicity, that $ b(\eta) \equiv c(\eta)$. 
Suppose now,
 that our initial state  is  weakly coupled (i.e.  $\phi_0$ sufficiently 
negative) and $({\cal H}_0, {\cal F}_0)$ are sufficiently small in string 
units. Even though ${\cal H}_0$ and ${\cal F}_0$
 are of the same order they do have different initial conditions. Let us 
therefore analyze what is the role of a magnetic field, small in string units.
The discussion  becomes simpler if we start looking at the $\eta$ behavior 
of the solutions. From Eqs. (\ref{s1}) we can argue that the evolution 
of $\phi'(\eta)$ is quite straightforward since 
for $\eta\rightarrow -\infty$ and 
$\phi'\rightarrow {\cal H}_0 - \Delta_0$ whereas for
 $\eta\rightarrow +\infty$ 
we have that $\phi'(\eta) \rightarrow {\cal H}_0 + \Delta_0$. Moreover, 
as we discussed in
the previous Section, $\phi''>0$ which implies that $\phi'$ is 
always an increasing function of $\eta$. As a consequence we will 
also have that ${\cal F}(\eta)$ will also be an increasing
function of $\eta$ whereas ${\cal H}$ is frozen to its constant 
value given by ${\cal H}_0$. According to  Eq. (\ref{c1}) the 
magnetic field intensity in string units determines,  together 
with $\phi_0$ the value of $\Delta_0$. Therefore, we 
conclude that the magnetic field intensity is essentially
responsible of the amount of growth of the dilaton energy
since it turns out that $|\phi'(+\infty) - \phi'(-\infty)|
= 2 \Delta_0$. Thus, by increasing the magnetic field
we observe an increase of the dilaton energy density. 
Notice, however, that this growth cannot be so large. 
Indeed, from Eq. (\ref{c1}) we notice that 
the relation between $\Delta_0$ and $B$ is suppressed 
by the second power of the (initial) coupling constant $g(\phi) 
=\exp(\phi/2)$. So, with our initial conditions, 
$\Delta_0$ is {\em small from the beginning} since $\phi_0$ is very 
negative. Notice that to have small coupling 
is essentially dictated by the form of our original action 
which cannot be trusted if $g(\phi)$ gets of order one.
This qualitative behavior of the solutions 
is illustrated in Fig. \ref{f3a}--\ref{f3b} 
in few specific cases. 
As a  a reference value 
for ${\cal H}_0$ and for $\phi_0$  we take  $0.001$ and 
$-5$ since we want to deal with small curvatures in String units
and small couplings.
In Fig. \ref{f3a} (left)
 we report the solutions in the case  of a quite large magnetic 
field $B\sim 1$. The dot-dashed line denotes the behavior of
${\cal H}$ (constant), the full (thin) line denotes the 
evolution of ${\cal  F}(\eta)$ whereas the behavior of 
the full (thick) line denotes the evolution of $\phi'$. 
It is interesting to compare this plot with the 
ones illustrated in Fig. \ref{f3b} where (left
plot) we report the solution (\ref{s1}) with the very same 
choice of initial conditions of Fig. (\ref{f3a}) 
but with different magnetic field, namely 
$B=0.001$ in String units. We see that the growth of 
$\phi'$ is reduced (by two order of magnitude). 

What about the evolution of the shear parameter?
Let us look at the left plots reported in 
Fig. \ref{f3a}, \ref{f3b}  where
the evolution of
\begin{equation}
r(\eta) = \frac{3 [ {\cal H}(\eta) - {\cal F}(\eta)]}{ [{\cal H}(\eta) 
+ 2 {\cal F}(\eta)] }
\end{equation}
is given for the same set 
of parameters we just discussed. 
In Fig. \ref{f3a} (where $B\sim 0.1$) we would 
say that the shear parameter decreases.
This is certainly true but if we look
at the numbers we can see that 
 $r(\eta)$ starts of order 
one and essentially remains of order one. 
In Fig. \ref{f3b} 
(i.e., $B=0.01$) 
$r(\eta)$ seem to decrease, but again, a more correct 
statement would be that $r(\eta)$ remains 
of order one (i.e. $r(+\infty) \sim - 0.88$), up to a transition period.
Thus, in spite of 
its  smallness,
 the magnetic field has always the property
 of {\em conserving}
the anisotropy. The negative sign in $r(\eta)$ 
is only due to the fact that ${\cal F}>{\cal H}$ for any 
$\eta$. Thus, as far as the tree-level evolution is 
concerned we can say  that no isotropization is 
observed as a consequence of the inclusion 
of the magnetic field. This situation should be contrasted with what 
happens when the string tension corrections are included.

The second effect we see is that the smaller is the 
value of the magnetic field, the milder is the transition to the vacuum 
solutions. We said, in the 
previous section that the vacuum solutions are essentially 
given by ${\cal H}= {\rm constant}$, 
${\cal  F}= constant$, ${\cal \phi}'={\rm constant}$.
For $\eta\rightarrow +\infty$ the magnetic solutions 
should then tend to  the vacuum solutions. Thus, the 
influence of the magnetic field is mainly in the transition period
from the initial state to the 
(asymptotic vacuum) solution. 

In order to show that the large $\eta$ behavior  of our solutions
is Kasner-like, let us look explicitly to the cosmic time 
description of the system.
Let us look at the evolution of our system in cosmic time. 
The relation between $\eta $ and $t$ can 
be obtained by inserting Eqs. (\ref{s1}) into Eq. (\ref{time})
whose integrated version becomes 
\begin{equation}
\frac{t}{ a_0 b_0^2 e^{\phi_0}} = e^{\phi_1} \frac{ e^{ 2 ( {\cal F}_0 
+ {\cal H}_0) \eta + \Delta_0\eta}}{[ 2 ( {\cal F}_0 + {\cal H}_0) + \Delta_0]}
+ \frac{ e^{ 2 ( {\cal F}_0 + {\cal H}_0) \eta 
- \Delta_0\eta}}{[ 2 ({\cal F}_0 + {\cal H}_0) - \Delta_0]}.
\label{concos}
\end{equation}
Notice that the only two independent quantities 
determining the relation between $\eta$ and $t$ are 
${\cal H}_0$ and $\Delta_0$ (which is related to $B$ according to 
Eq. (\ref{c1})) since ${\cal F}_0$ can be expressed   in terms of $\Delta_0$
and ${\cal H}_0$.

If we now take the large $\eta$ behavior of Eq. (\ref{concos})
 and we substitute 
back into Eqs. (\ref{s1}) we do find that the two scale factors 
can be expressed, 
in cosmic time, as 
\begin{equation}
a(t) \sim t^{\alpha},~~~b(t) \sim t^{\beta},
\end{equation}
where 
\begin{equation}
\alpha= \frac{ {\cal H}_0 }{2( {\cal H}_0 + {\cal F}_0) + \Delta_0},~~~
\beta = \frac{ {\cal F}_0 + {\cal H}_0 
+ \Delta_0}{2 ( {\cal H}_0 + {\cal F}_0 ) + \Delta_0}.
\end{equation}
We can notice that $\alpha^2 + 2 \beta^2 =1$. This can be easily
seen by recalling the definition of 
$\Delta_0^2 = {\cal H}_0^2 + 2(2 {\cal H}_0 {\cal F}_0 + {\cal F}_0^2)$
which can be deduced from Eqs. (\ref{c1}) and (\ref{eqtrial}).

\begin{figure}
\begin{center}
\begin{tabular}{|c|c|}
      \hline
      \hbox{\epsfxsize = 6.5 cm  \epsffile{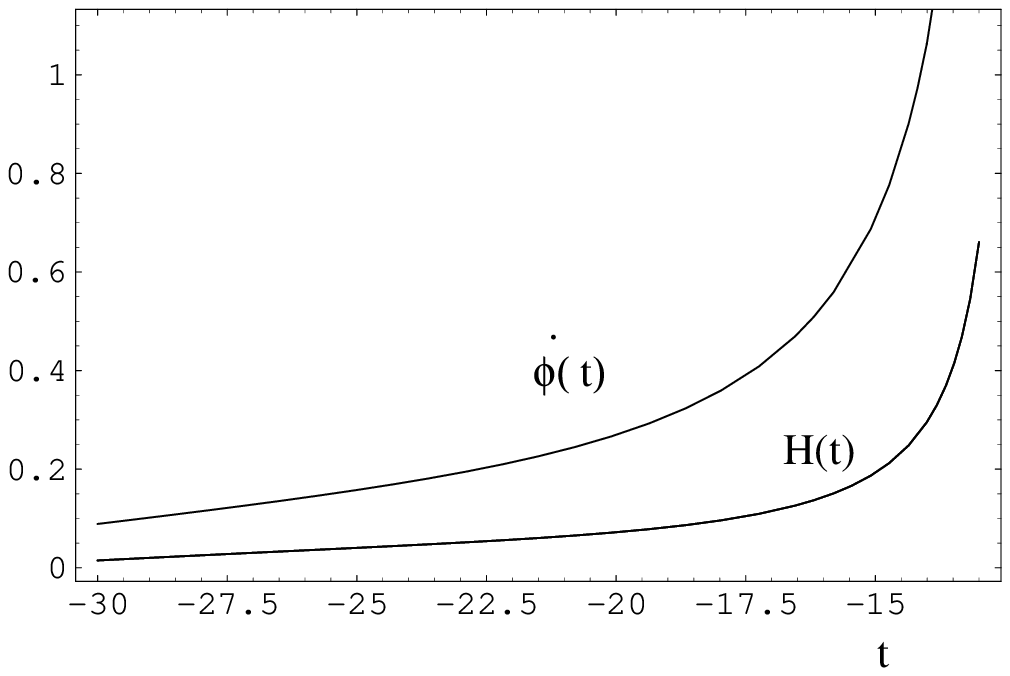}} &
      \hbox{\epsfxsize = 6.5 cm  \epsffile{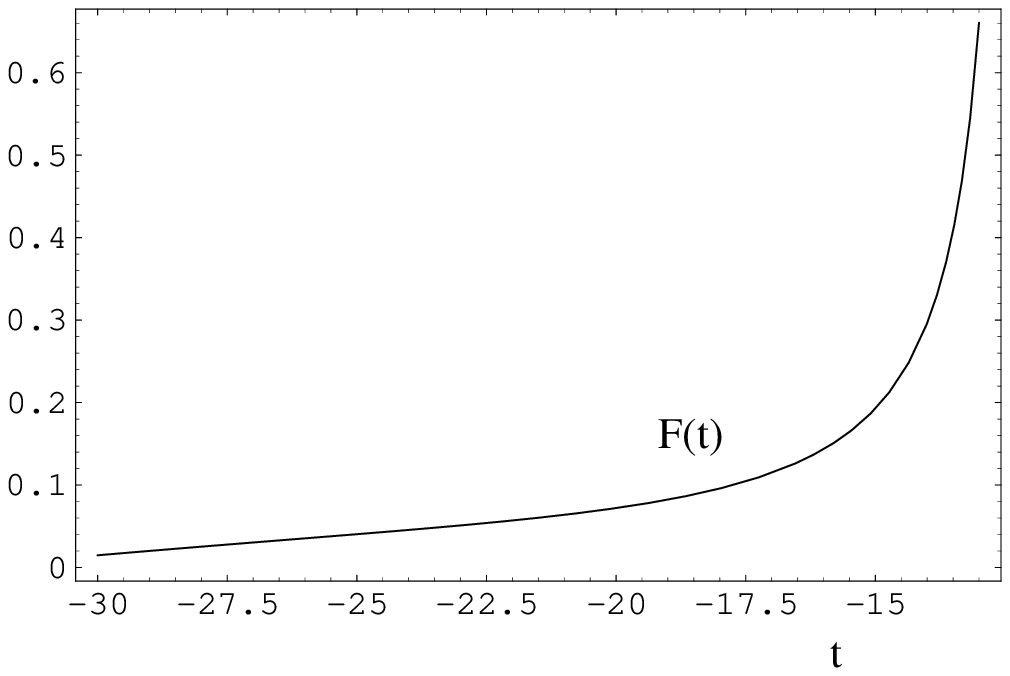}} \\
      \hline
\end{tabular}
\end{center}
\caption[a]{We illustrate the behavior of $H(t)$, $\dot{\phi}$ (left plot)
 and of 
$F(t)$ (right plot) in the case of a magnetic field of the order 
of $B\sim 0.1$ in String units.}
\label{f3e}
\end{figure}

\begin{figure}
\begin{center}
\begin{tabular}{|c|c|}
      \hline
      \hbox{\epsfxsize = 6.5 cm  \epsffile{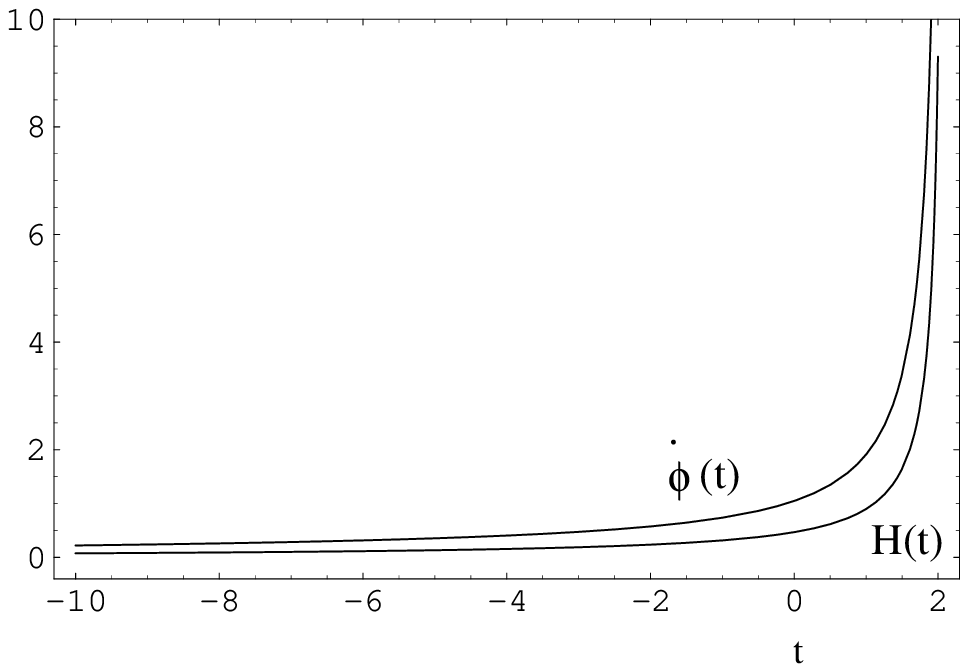}} &
      \hbox{\epsfxsize = 6.5 cm  \epsffile{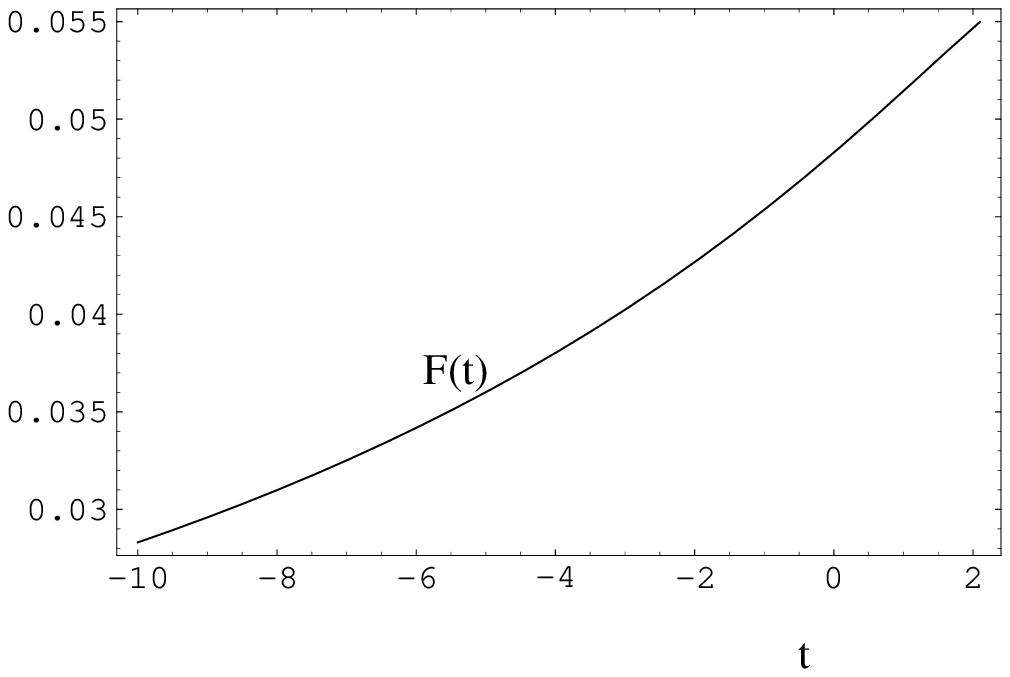}} \\
      \hline
\end{tabular}
\end{center}
\caption[a]{ We illustrate the behavior of $H(t)$, $\dot{\phi}$ (left plot)
 and of 
$F(t)$ (right plot) in the case of a magnetic field of the order 
of $B\sim 0.1$ in String units.}
\label{f3d}
\end{figure}

So, close to the singularity, for $\eta\rightarrow\infty$
the solutions are Kasner-like.
Notice that the presence of the vector field (with the specific
 four-dimensional coupling required by the low energy effective action) 
does not lead to any type of oscillatory stage.  
This is evolution is in sharp contrast with one obtains in the case 
of a vector field (included following the Kaluza-Klein ansatz) in the Einstein
frame \cite{11}. The reason for this difference stems  from from the difference
 in the couplings.
In Fig. \ref{f3e} the evolution of $\dot{\phi}$, $H(t)$ and $F(t)$ 
are illustrated in the case of $B\sim 0.1$. Notice that $t\rightarrow 0$
does correspond to the large $\eta$ behavior. The singularity  is located, 
in this example for $t\sim 2$. 
In Fig. \ref{f3d} we report the same solution illustrated in Fig.
\ref{f3e} but with a smaller magnetic field, $B=0.01$. As one can expect
the singularity and the vacuum limit is reached before than in the 
previous case.

\renewcommand{\theequation}{4.\arabic{equation}}
\setcounter{equation}{0}
\section{Magnetic Solutions and the High Curvature Regime}

In the previous two Sections we examined fully anisotropic (magnetic) 
cosmologies using the tree level action. In this Section we are going 
to extend and complement our results with the addition to the tree-level 
action of the first string tension correction. In the presence
of the first $\alpha'$ correction the action becomes 
\begin{equation}
S= - \frac{1}{2 \lambda^2_{s}}\int d^4x \sqrt{- g} e^{- \phi} \biggl[ R
+ g^{\alpha\beta} \partial_{\alpha}\phi \partial_{\beta} \phi 
+\frac{1}{4}F_{\alpha\beta}F^{\alpha\beta} - \frac{
\omega\lambda_{s}^2}{4}\biggl(R_{GB}^2 - (g^{\alpha\beta} 
\partial_{\alpha}\phi\partial_{\beta}\phi)^2\biggr)\biggr],
\label{action4}
\end{equation}
where $R_{GB}^2$ is the Gauss-Bonnet invariant expressed in
terms of the Riemann, Ricci and scalar curvature invariants
\begin{equation}
R_{GB}^2 = R_{\mu\nu\alpha\beta}R^{\mu\nu\alpha\beta} - 4
R_{\mu\nu}R^{\mu\nu} +R^2,
\end{equation}
and $\omega$ is a
numerical constant of order 1 which can be precisely computed
depending upon the specific theory we deal with (for instance 
$\omega=-1/8$ for heterotic strings). 
In principle the magnetic field should also appear in the corrections to the
tree-level action. However, it turns out that, the terms which would be 
significant for the type of configurations examined in this paper, the gauge 
fields appear beyond the first $\alpha'$ correction. For example terms like 
$F^{\alpha\beta}F^{\mu\nu} R_{\alpha\beta\mu\nu}$ will appear 
to higher order in $\alpha'$. Moreover, possible terms  like 
$F^{\alpha\beta} R_{\alpha\beta}$ are vanishing in the case of constant and 
homogeneous magnetic field directed, say, along the $x$ direction. Finally, 
other possibly relevant terms (involving contractions of the vector potentials
with the Ricci or Riemann tensors) are forbidden by gauge invariance.

Of course, the spirit of our analysis of the high curvature regime
 is only semi-quantitative. After all, the first curvature correction can  be 
thought as illustrative, since, ultimately {\em all} the $\alpha'$ corrections
will turn out to be important. The only way of turning-off the possible 
effect of curvature corrections is to have {\em explicit} solutions
whose curvature invariants are already regularized at tree level like 
in the examples presented in \cite{12} where it was shown that there 
exist weakly inhomogeneous 
solutions of the tree level action which are regular and geodesically  
complete without the addition of any curvature or loop corrections to the 
tree level action. If this is not the case, the curvature corrections have 
to be certainly included. The inclusion of higher order curvature corrections
 can be studied either in the Einstein frame \cite{12b} or in the String frame 
\cite{12c}. Another interesting approach has been outlined in \cite{12d}.

\begin{figure}
\begin{center}
\begin{tabular}{|c|c|}
      \hline
      \hbox{\epsfxsize = 6.5 cm  \epsffile{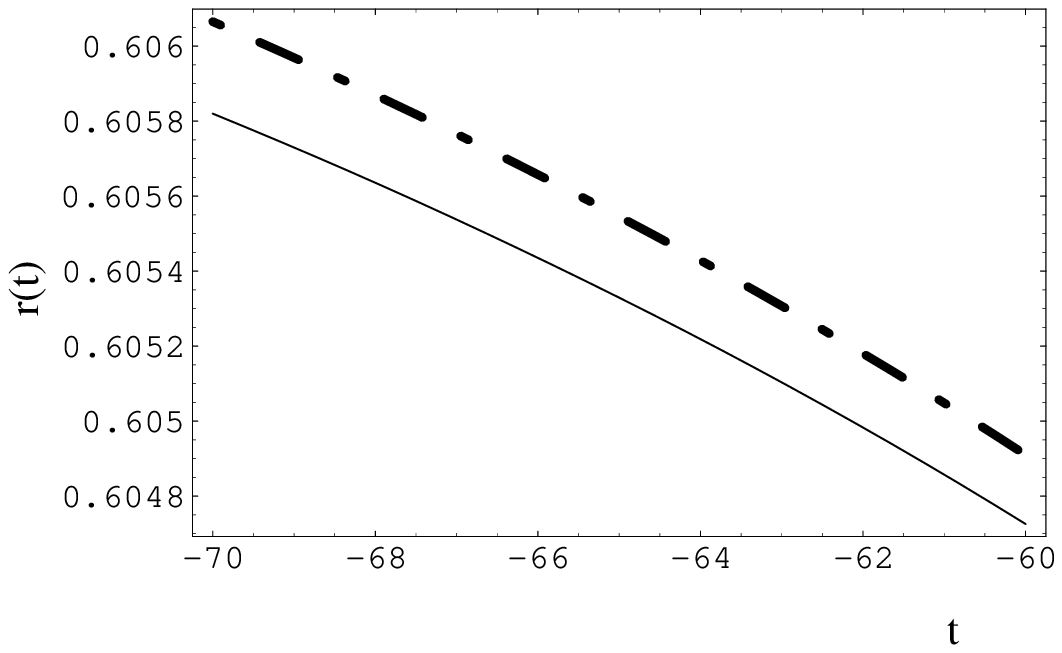}} &
      \hbox{\epsfxsize = 6.5 cm  \epsffile{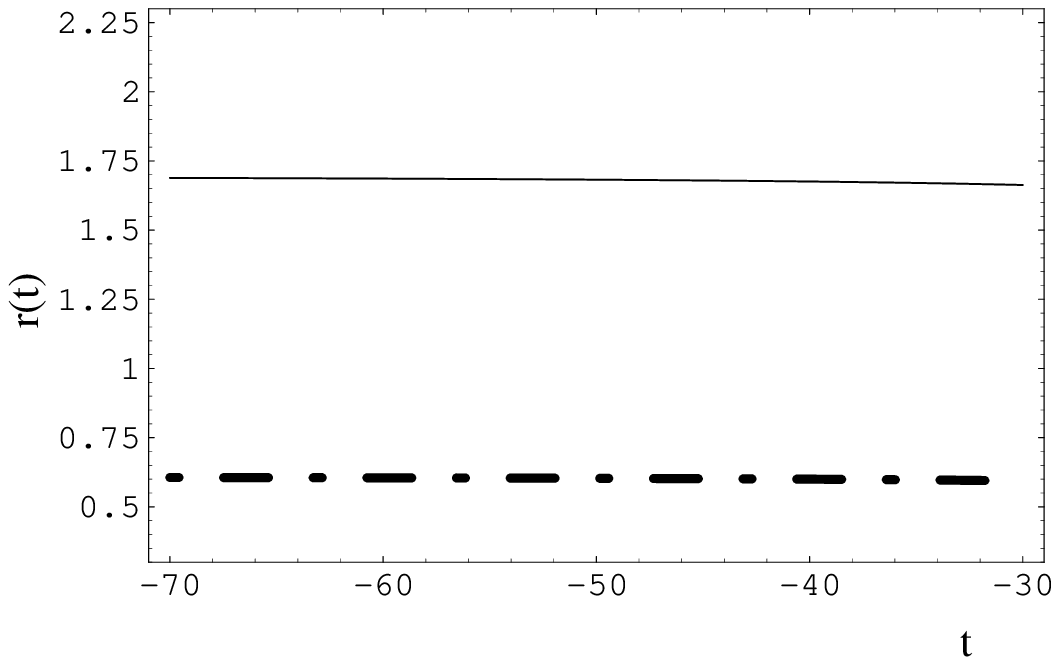}} \\
      \hline
\end{tabular}
\end{center}
\caption[a]{ We illustrate the numerical solutions of the 
equations of motions {\em with } string tension corrections and 
we plot the shear parameter.
We select expanding initial conditions (in the String frame).
In this particular example (left plot) we choose the magnetic field
to  be  $B=0$ (dot dashed line), and $B=0.01$ (full line). 
The difference between 
the two cases is quite minute and in order to show it
we do not plot the full evolution of $r(t)$ (which goes to zero for 
$t>0$). At the right 
we choose  $B= 0.01$ and we change the initial condition within
the vacuum solutions. The dot dashed line is for $b(t) \sim t^{-4/9}$ 
and the full line for $b(t)\sim t^{-0.2}$. Also in this second case 
the difference is quite minute. We decided to plot only a limited range of 
time steps in  order to stress the numerical difference between 
different values of the initial shear and of the primordial magnetic field.
The full picture looks like the ones reported in Fig. \ref{f7}. }
\label{f6}
\end{figure}

\begin{figure}
\begin{center}
\begin{tabular}{|c|c|}
      \hline
      \hbox{\epsfxsize = 6.5 cm  \epsffile{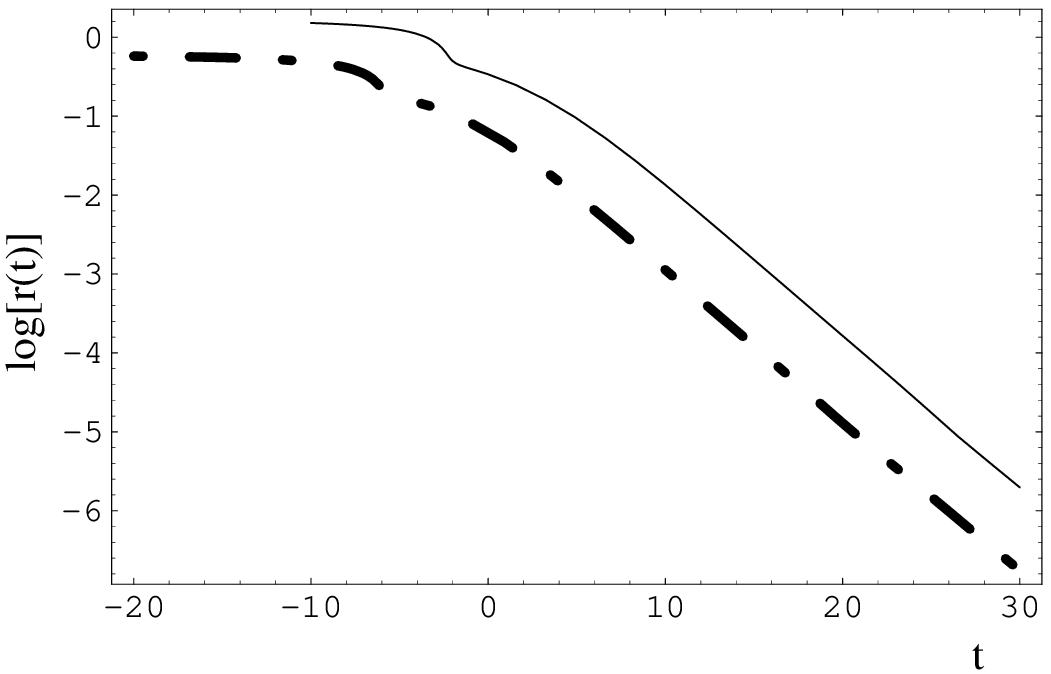}} &
      \hbox{\epsfxsize = 6.5 cm  \epsffile{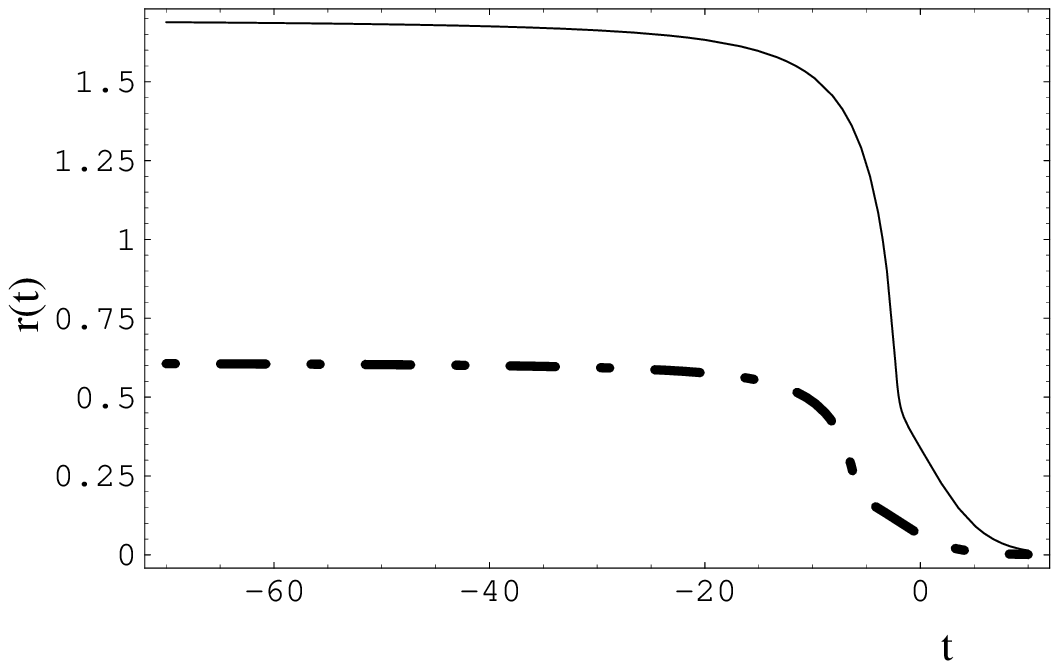}} \\
      \hline
\end{tabular}
\end{center}
\caption[a]{We report the evolution of the shear parameter in the String phase.
With the dot-dashed line we have the case 
$B=0$, with the full line the case $ B=0.01$. If a magnetic field is present 
the shear parameter is larger than in the case of zero magnetic field. 
It is of crucial importance to notice, for our purposes, 
that the shear parameter 
does not tend towards a small (finite value) but it decreases towards $0$. 
In order to
stress this aspect we also plot the logarithm (in ten basis) of the modulus 
of the shear parameter. We can see, as previously discussed, that 
the duration of the string phase is proportional to the drop in $r(t)$. }
\label{f7}
\end{figure}

For a reasonably swift derivation of the equations of 
motion it is useful to write the action in terms of the relevant degrees 
of freedom. In order to get correctly the constraint equation (i.e. the 
$(00)$ component of the beta functions) it is appropriate 
to keep the lapse function $N(t)$ it its general form
\begin{equation}
g_{\mu\nu} = {\rm diag}[ N(t)^2, -e^{2 \alpha(t)}, - e^{2 \beta(t)}, 
- e^{2 \beta(t)}],
\label{N}
\end{equation}
where we parameterized the two scale factors with an exponential notation.
Only {\em after} the 
equations of motion have been derived we will set $N(t)=1$ corresponding 
to the synchronous time gauge. In the metric (\ref{N}) the previous action 
(\ref{action4}) becomes, after integration by parts,
\begin{equation}
S= \frac{1}{2 \lambda_{s}^2} \int d t e^{\alpha + 2 \beta
-\phi}\biggl\{ \frac{1}{N} \biggl[ - \dot{\phi}^2 - 2 \dot{\beta}^2 -
4 \dot{\alpha} \dot{\beta} + 2 \dot{\alpha} \dot{\phi} + 4 \dot{\beta}
\dot{\phi}-\frac{1}{2} B^2 e^{- 4 \beta(t)}
\biggr] + \frac{\omega\lambda_{s}^2}{4 N^3}\biggl[ 8 \dot{\phi}~
\dot{\alpha}~ \dot{\beta}^2 - \dot{\phi}^4\biggr]\biggr\}.
\label{action5}
\end{equation}
By varying Eq. (\ref{action5}) with
respect to the lapse function $N(t)$ and imposing, afterwards, the
cosmic time gauge we get the constraint
\begin{equation}
\dot{\phi}^2 + 2 \dot{\beta}^2 + 4 \dot{\alpha}\dot{\beta} 
- 2 \dot{\alpha}\dot{\phi} - 4 \dot{\beta} \dot{\phi} 
-\frac{1}{2}B^2 e^{-4 \beta}+
\biggl[\frac{3}{4} \dot{\phi}^4 - 6 \dot{\alpha} \dot{\phi} 
\dot{\beta}^2\biggr]=0,
\label{004}
\end{equation}
where we took string units $\lambda_s=1$ and $\omega=1$.
By varying the action with respect to $\alpha$, $\beta$ and $\phi$ we
get, for $N(t)=1$ the diagonal components of the beta functions and 
the dilaton equation,
\begin{eqnarray}
&& 4 \ddot{\beta} - 2 \ddot{\phi} + (4 \dot{\beta} - 2 \dot{\phi})
(\dot{\alpha} + 2 \dot{\beta} - \dot{\phi}) - 2 \biggl[ (\dot{\alpha} +
2 \dot{\beta} - \dot{\phi})\dot{\phi} \dot{\beta}^2 
+ \ddot{\phi} \dot{\beta}^2+ 2
\dot{\beta} \ddot{\beta}\dot{\phi}\biggr] +
L(t)=0,
\\
&& 4 (\ddot{\beta} + \ddot{\alpha} - \ddot{\phi}) + 2 B^2 e^{- 4 \beta} 
+ 4( \dot{\alpha} +
\dot{\beta} - \dot{\phi})( \dot{\alpha} + 2 \dot{\beta} -\dot{\phi}) -
4 \biggl[ \ddot{\beta}\dot{\alpha} \dot{\phi} + \dot{\beta}
\ddot{\alpha} \dot{\phi} + \dot{\beta} \dot{\alpha} \ddot{\phi} + 
\dot{\beta} \dot{\alpha} \dot{\phi} (\dot{\alpha} + 2 \dot{\beta} -
\dot{\phi})\biggr] + 2 L(t)=0,
\\
&&
2 (\ddot{\phi} - \ddot{\alpha} - 2 \ddot{\beta}) + 2 (\dot{\phi} -
\dot{\alpha} -2 \dot{\beta})(\dot{\alpha} + 2 \dot{\beta} -
\dot{\phi}) -\biggl[   2 \dot{\beta}(\ddot{\alpha}\dot{\beta} + 2 \dot{\alpha}
\ddot{\beta}) + (2 \dot{\alpha} \dot{\beta}^2 -
\dot{\phi}^3)(\dot{\alpha} + 2 \dot{\beta} - \dot{\phi}) - 3
\dot{\phi}^2 \ddot{\phi} \biggr] -L(t)
=0,
\label{alpha}
\end{eqnarray}
where we defined 
\begin{equation}
L(t) = -\dot{\phi}^2 -2 \dot{\beta}^2 - 4 \dot{\alpha}\dot{\beta} + 2
\dot{\alpha} \dot{\phi} + 4 \dot{\beta}\dot{\phi} - \frac{1}{2} B^2  
e^{- 4 \beta}+ \frac{1}{4}(8
\dot{\phi}\dot{\alpha} \dot{\beta}^2 - \dot{\phi}^4).
\end{equation}
We can numerically integrate this system. The technique is very similar 
to the one used in the case of fully anisotropic string cosmologies 
with zero magnetic field \cite{6,12c}. Our results are illustrated in 
Fig. \ref{f6} and \ref{f7}. There are two physically different cases.

\begin{figure}
\begin{center}
\begin{tabular}{|c|c|}
      \hline
      \hbox{\epsfxsize = 6.5 cm  \epsffile{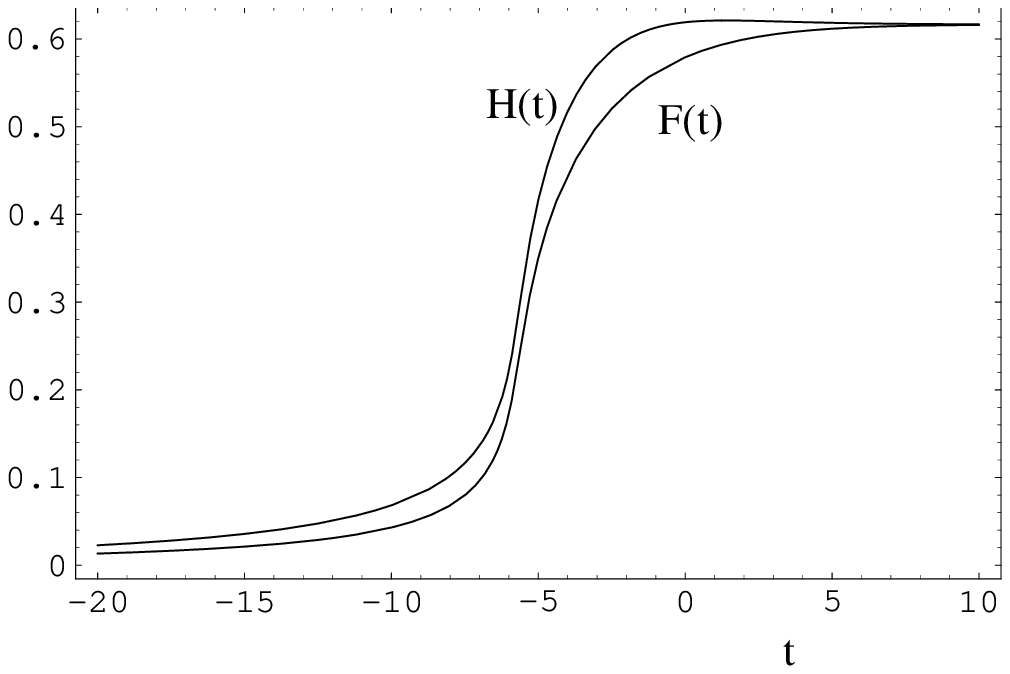}} &
      \hbox{\epsfxsize = 6.5 cm  \epsffile{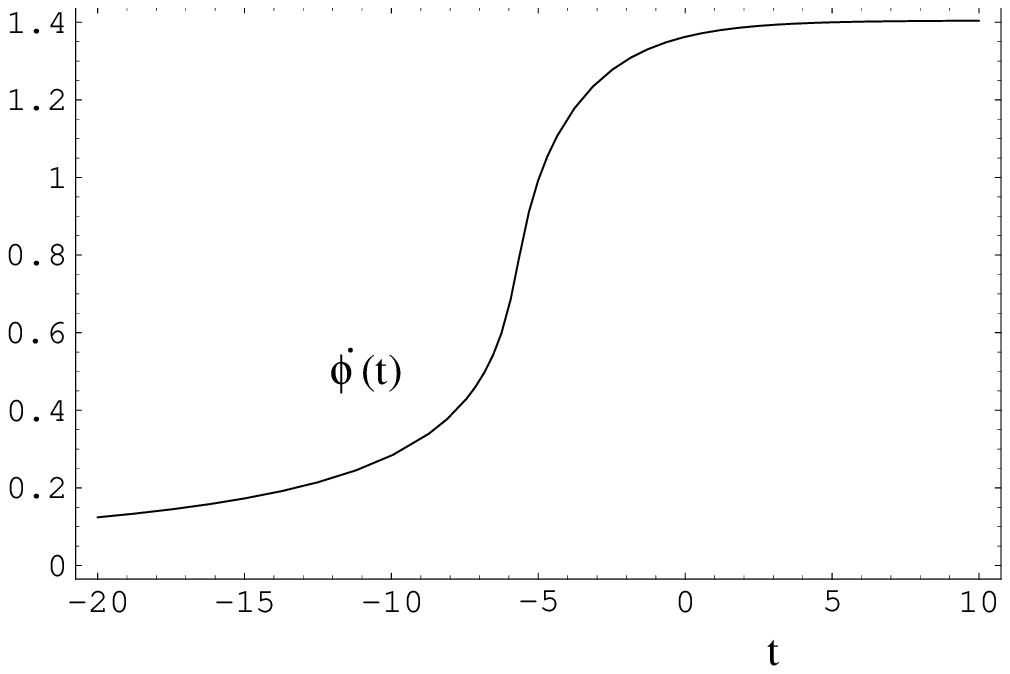}} \\
      \hline
\end{tabular}
\end{center}
\caption[a]{We report the result of the numerical integration of 
Eqs. (\ref{alpha}) in the case where the initial conditions are 
mildly anisotropic. We choose $B=0.01$. }
\label{f6a}
\end{figure}
We integrate this system by imposing, as initial conditions 
small curvature and small dilaton coupling. We also impose  mildly 
anisotropic initial conditions (corresponding to 
$\dot{a}>0$, $\dot{b}>0$ and $\ddot{a}>0$, $\ddot{b}>0$).
One of the results stressed in the previous Section is that the tree-level 
solutions with magnetic field are singular. Moreover, close to the 
singularity no oscillatory behavior is present. Therefore, as time goes by and
as the solutions approach the singularity the vacuum solutions are recovered.
One can then argue that if the $\alpha'$ corrections become relevant when 
the vacuum regime has been already recovered, nothing should change 
with respect to the case where $B=0$.  This in some sense is what happens, but
not exactly.
If the magnetic field is large in string units (i.e. $B> 1$) this system 
evolves towards a singularity, as expected. In this case, it is 
certainly true that the tree level solutions evolve towards their vacuum 
limit. However, in this limit the solutions get more and more anisotropic. So
the hypothesis $B>1$ simply contradicts 
the assumption  of mild anisotropy.  
From our point of view it is more interesting the case with $B<1$. 
In this case mildly anisotropic initial conditions will be attracted 
towards an isotropic fixed point. One can understand this looking at 
Eqs. (\ref{alpha}). For $t\rightarrow 0$ $b(t)$  is  
monotonically increasing (if we select, as we do, 
expanding initial conditions). Now the magnetic field is always suppressed by
$b^{-4}(t)= e^{- 4\beta(t)}$ terms in Eqs. (\ref{alpha}). Therefore, 
an initially small magnetic field will become more and more sub-leading. 
The solutions will then be attracted towards the quasi-isotropic fixed point
$\dot{\alpha}\sim \dot{\beta} \sim 0.606$ and $\dot{\phi} \sim 1.414$. 
In this case the curvature invariants are finite everywhere.
We say that the fixed point is quasi-isotropic since the amount of anisotropy
depends upon the duration of the high curvature scale. For example, as 
illustrated in Fig. \ref{f7}, it can well happen that for a very short 
stringy phase the amount of anisotropy measured by the shear parameter 
$r(t)$  will be of the order of $10^{-3}$. 

\renewcommand{\theequation}{5.\arabic{equation}}
\setcounter{equation}{0}
\section{The Fate of the Anisotropies in the Post-Big-Bang evolution}

In this Section we would like to investigate the fate of the anisotropies
possibly present in the dilaton-driven and in the String phase. In particular
we would like to understand in which limit these anisotropies can be either 
reduced to an acceptable value or completely washed out. 
For numerical purposes  we find useful to exploit, in the present Section, 
 the Einstein frame picture. 
The evolution equations in the Einstein frame are obtained in Appendix A. 
By linearly combining Eqs. (\ref{E00}), (\ref{Exx}) 
and (\ref{Eyy}) of Appendix A we obtain a more readable 
form of the system, namely
\begin{eqnarray}
&&\dot{H} + H( H + 2 F) = \frac{\rho}{6} - \frac{w}{2} e^{-\phi},
\label{I}\\
&&\dot{F} + F ( H + 2 F)  = \frac{\rho}{6} + \frac{w}{2} e^{-\phi}, 
\label{II}\\
&&\ddot{\phi} + ( H + 2 F) \dot{\phi} = w e^{-\phi} ,
\label{III}\\
&&(H + 2 F)^2 - (H^2 + 2 F^2) = \frac{{\dot{\phi}}^2}{2} 
+ w e^{- \phi} + \rho,
\label{IV}\\
&&\dot{\rho} + \frac{4}{3}( H + 2 F) \rho =0, ~~~\dot{w} + 4 F w=0,~~~~
w= \frac{B^2}{2 b^4},
\label{V}
\end{eqnarray}
where we assumed the presence of a radiation fluid $p= \rho/3$ accounting
for the field modes excited, via gravitational instability, 
during the dilaton driven phase \cite{13} and re-entering in the post-big-bang 
phase. It is in fact well known that the ultraviolet modes of a field 
parametrically amplified behave as a radiation fluid \cite{14}.

From  Eqs. (\ref{I})--(\ref{V}) we can directly obtain the evolution equations
for the quantities we are interested in, namely
\begin{equation}
r(t) = \frac{3(H - F)}{ H + 2 F}, ~~~n(t) = \frac{ H + 2 F}{3}, ~~~~ 
q(t) = \frac{w(t)}{\rho(t)},
\label{definitions}
\end{equation}
where $n(t)$ is the mean expansion parameter $r(t)$ is the shear anisotropy
parameter and $q(t)$ the fraction of magnetic energy in units of radiation 
energy. Notice that $q(t)$ {\em does not} correspond to the critical
fraction of magnetic energy density since, in principle, we have to take into
account the energy density of the dilaton field. By subtracting Eq. (\ref{II}) 
from Eq. (\ref{I}) and by using the constraint (\ref{IV}) together with the 
definitions (\ref{definitions})  we get the 
shear evolution. By summing Eqs. (\ref{I}) and (\ref{II}) with similar 
manipulations we get the evolution of $n$. The resulting system, equivalent 
to the one of Eqs. (\ref{I})--(\ref{V}), but directly expressed in terms 
of $n$, $r$ and  $q$ reads 
\begin{eqnarray}
&&\dot{n} r + \dot{r} n + 3 n^2 r = - q \rho e^{-\phi},
\label{AI}\\
&& 3 \dot{n} + 9 n^2 = \frac{\rho}{2} ( 1 + q e^{- \phi}),~~~ \dot{q} 
- \frac{4}{3}n r q =0 ,
\label{AII}\\
&& 6 n^2 ( 1 - \frac{r^2}{9} ) = \frac{{\dot{\phi}}^2}{2} + 
\rho ( 1 + q e^{-\phi}), ~~~ \dot{\rho} +  4 n\rho =0,~~~ \ddot{\phi} +
 3 n \dot{\phi}  = \rho q e^{- \phi}.
\label{AIII}
\end{eqnarray}

\begin{figure}
\begin{center}
\begin{tabular}{|c|c|}
      \hline
      \hbox{\epsfxsize = 6.5 cm  \epsffile{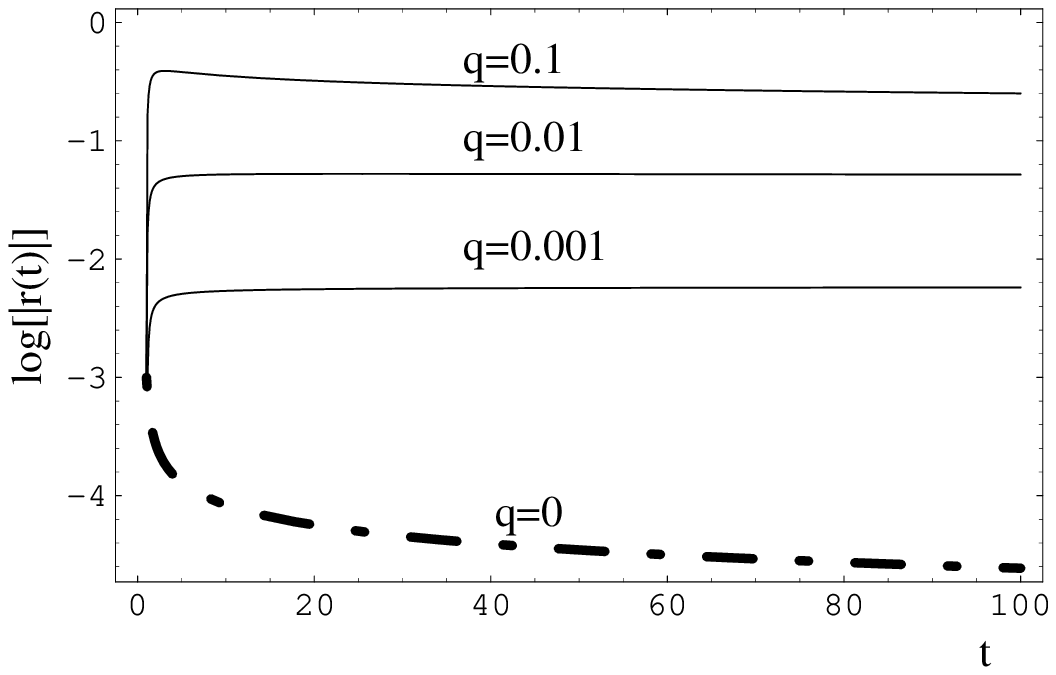}} &
      \hbox{\epsfxsize = 6.5 cm  \epsffile{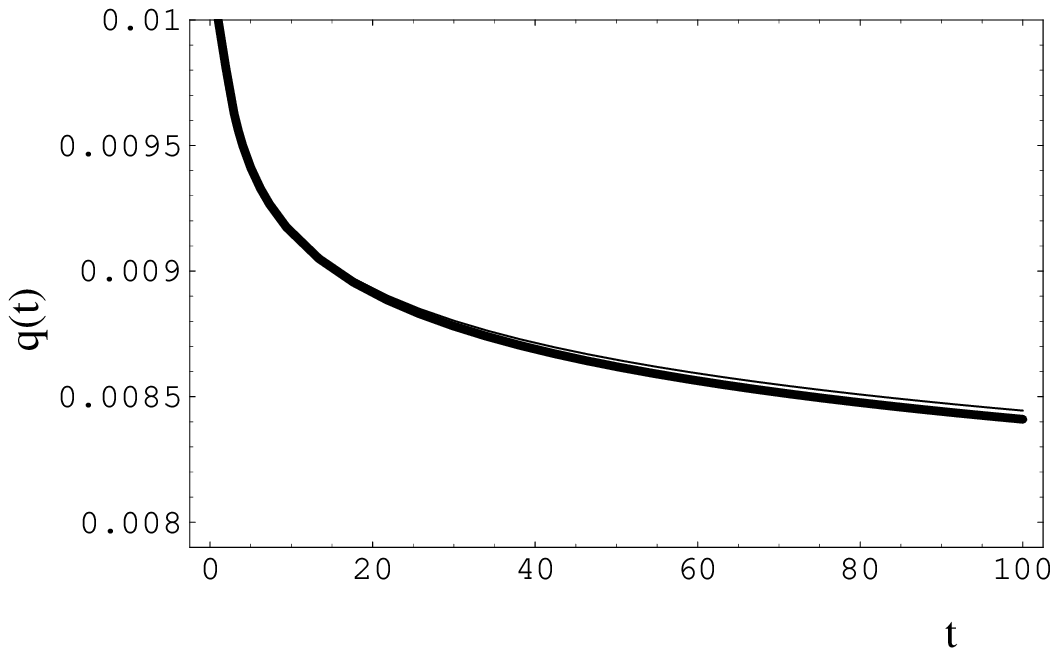}} \\
      \hline
\end{tabular}
\end{center}
\caption[a]{We report the results of
 the numerical integration of Eqs. (\ref{BI})--(\ref{BIII}). We take 
$r(t_0)\simeq 10^{-3}$ and we integrate the system forward in time for 
different 
values of $q(t_0)$. In the left plot we illustrate the evolution of the 
logarithm (in ten basis) of the modulus of the shear parameter. With 
the dot-dashed (thick) line we represent the case $q=0$ where the shear 
parameter is known to decrease sharply as $t^{-1/2}$. If, initially,
 the critical balance between the magnetic energy density and 
the radiation energy density increases, then $r$ gets attracted towards and
 asymptotic value as described by Eqs. (\ref{shearrad}). At the right
(full thick line) we report 
the evolution of $q$ for the case $q(t_0) = 0.01$. With the thin line 
we report 
the qualitative estimate based on Eqs. (\ref{shearrad}) and obtained
 by solving, 
approximately, Eqs. (\ref{BI})--(\ref{BIII}) for $r<1$ and $q<1$.}
\label{f8}
\end{figure}

We want now to study the combined action of the dilaton and of
a radiation fluid in a post-big-bang phase.
Let us start from the case of medium size anisotropies. Suppose, in other 
words that after a string phase of intermediate duration the shear parameter is
of the order of $10^{-3}$. As we saw from the previous Section this value is 
not excluded. Before analyzing the general case let us recall the results 
of the case where the dilaton is absent. In this case the system of Eqs. 
(\ref{AI})--(\ref{AIII}) can be simplified as follows 
\begin{eqnarray}
&&\dot{n} r + \dot{r} n + 3 n^2 r = - q \rho ,
\label{BI}\\
&& 3 \dot{n} + 9 n^2 = \frac{\rho}{2} ( 1 + q),~~~ \dot{q} 
- \frac{4}{3}n r q =0 ,
\label{BII}\\
&& 6 n^2 ( 1 - \frac{r^2}{9} ) = 
\rho ( 1 + q e^{-\phi}), ~~~ \dot{\rho} +  4 n\rho =0.~~
\label{BIII}
\end{eqnarray}
We are interested in the case of magnetic fields which are under-crtical 
, namely $q(t_0)<1$ where $t_0$ is the initial integration time. Suppose,
moreover, that $r(t_0) \sim 10^{-3}$. Needless to say that 
this is just an illustrative value for $r$ and {\em it is not meant to be 
of any particular theoretical relevance}. As we stated clearly in the 
previous Section (see for instance Fig. \ref{f6} ) if the string phase is long
the initial shear parameter can be much smaller. 
The results of the numerical integration 
are summarized in Fig. \ref{f8} where we report the logarithm (in ten basis)
of the modulus of the shear parameter.
At the bottom of the left plot,
 with the dot-dashed line, is reported the case of zero magnetic 
field (i.e. $q(t_0) =0$). This is the simplest case since we can approximately 
solve the system of Eqs. (\ref{BI})--(\ref{BIII}) in the limit 
of small $r$. From Eq. (\ref{BI}) we can find that a consistent solution 
is 
\begin{equation}
\dot{r} \simeq -\frac{r}{2 t} - \frac{3 q}{  t} 
, ~~~\dot{q} \simeq \frac{2}{3 t} r q.
\label{shearrad}
\end{equation}
So, if $q=0$ we can clearly see that $r(t) \sim t^{-1/2}$. This is nothing 
but the well known result that in a radiation dominated phase the 
shear parameter decreases as $1/\sqrt{t}$ and it is precisely what we 
find, numerically, in the dot-dashed line of Fig. \ref{f8} (left plot). 
If we switch-on the
magnetic field we also know, from Eqs. (\ref{shearrad}) that the anisotropy
will not decrease forever but it will reach an asymptotic value which 
crucially depends upon the balance between the magnetic energy density and
the radiation energy density. In fact from eqs. (\ref{shearrad}) we see 
that $\dot{r}=0$ for $ |r(t)|\rightarrow  6 |q(t)|$ and this is precisely 
what we observe in the full lines of Fig. \ref{f8} where the integration of the
Eqs. (\ref{BI})--(\ref{BIII}) is reported for different (initial) values of q. 
We see that the asymptotic value of attraction for $r$ is roughly six times
the initial value of $q$. The evolution of $q(t)$ when $\dot{r}(t) 
\rightarrow 0$
can be simply obtained by integrating once the second of Eqs. (\ref{shearrad})
for the case $r\sim - 6 q$. The result is that, taking for instance 
$q(t_0) =0.01$,
$q(t) \sim 1/\{ 4 \log[(t/t_0)] + 100\}$. This simple curve
 is reported in Fig. 
\ref{f8} at the right (full thin line). As we can see the full 
thick line (result
of the numerical integrations is practically indistinguishable).
 
The conclusion we draw for our specific case is very simple. If the 
string phase is not too long and if sizable anisotropies 
(i.e.  $r\sim 10^{-3}$) are still present at the end of the stringy phase, then
, in the approximation of a sudden ``freeze-out''
 of the dilaton coupling,
the left over of the primordial anisotropy will be washed out {\em provided }
the magnetic field is completely absent. If a tiny magnetic seed directed 
along the same direction of the anisotropy is present, then in spite
of the primordial anisotropy, the shear parameter will be attracted to a 
constant value essentially fixed by the size of the seed. 

\begin{figure}
\centerline{\epsfxsize = 7 cm  \epsffile{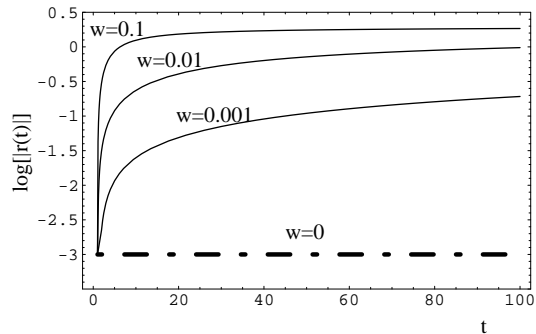}} 
\caption[a]{We report the resluts of the numerical integration of Eqs. 
(\ref{CI})--(\ref{CIII}) for different initial values of the magnetic 
energy density
in Planck units. As we can see (dot-dashed line) the shear parameter remains 
fixed to its initial value if the magnetic field is completely absent. As soon 
as the magnetic field increases the shear parameter gets attracted towards
 a fixed
point whose typical anisotropy is of order one. As in the previous cases 
we took $r(t_0)\sim 10^{-3}$ and we also took $\phi(t_0)\sim -0.1$. }
\label{f9}
\end{figure}

Let us now  investigate the opposite case. Let us imagine of switching-off 
completely the radiation background. Then the only non trivial 
variables determining the shear parameter and the mean expansion will be 
the dilaton coupling and the magnetic energy density.
In this case Eqs. (\ref{AI})--(\ref{AIII}) do simplify as follows
\begin{eqnarray}
&&\dot{n} r + \dot{r} n + 3 n^2 r = - w e^{-\phi} ,
\label{CI}\\
&& 3 \dot{n} + 9 n^2 = \frac{w}{2} e^{- \phi},~~~ \dot{w} 
+ 4( n - \frac{n r}{3})w =0,
\label{CII}\\
&& 6 n^2 ( 1 - \frac{r^2}{9} ) = \frac{{\dot{\phi}}^2}{2} + 
w e^{-\phi}, ~~~ \ddot{\phi} +3 n \dot{\phi}  = \rho q e^{- \phi}. 
\label{CIII}
\end{eqnarray}
As in the previous case  let us assume, in order to fix our ideas 
and in order to have a swift comparison, that 
$r(t_0)\sim 10^{-3}$.  The results of our analysis are reported in 
Fig. \ref{f9}.
Suppose first of all that the magnetic field is switched-off, then, 
as we see from the 
dot-dashed line the anisotropy is completely conserved. The introduction 
of a tiny magnetic seed will eventually make the situation even worse 
in the sense that the anisotropy will eventually grow to a constant value 
which depends on the magnetic field intensity. In Fig. \ref{f9} 
we see (full lines) that already with a magnetic energy density of the 
order of $10^{-3}$ (in Planck units) the  increase in the anisotropy cannot 
be neglected. Therefore, if the transition from the string phase to the 
radiation dominated phase does occur through an intermediate dilaton 
dominated phase of decreasing coupling we have to accept that  the 
anisotropy will be conserved {\em provided} the magnetic field is 
absent. If a magnetic field is present the anisotropy gets gets frozen to a 
constant value which can be non negligible depending upon the size of 
the magnetic seed. In very rough terms our analysis seems to disfavor 
mildly anisotropic models with a short string phase and with the simultaneous
occurrence of a long dilaton dominated phase prior to the onset of the 
radiation epoch. So the combination of {\em short} string phase, {\em long}
dilaton driven phase and  radiation phase {\em shorter than usual} might 
give too large shear, say, prior to nucleosynthesis. Our conclusion can be 
even more problematic if it is present a sizable magnetic field. 

\begin{figure}
\begin{center}
\begin{tabular}{|c|c|}
      \hline
      \hbox{\epsfxsize = 6.5 cm  \epsffile{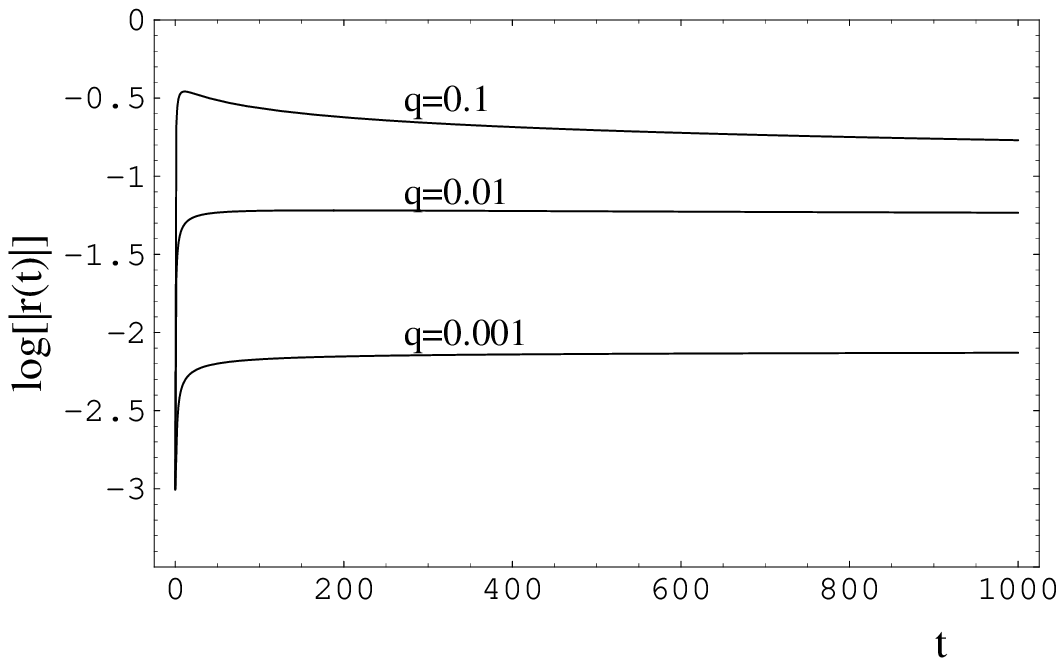}} &
      \hbox{\epsfxsize = 6.5 cm  \epsffile{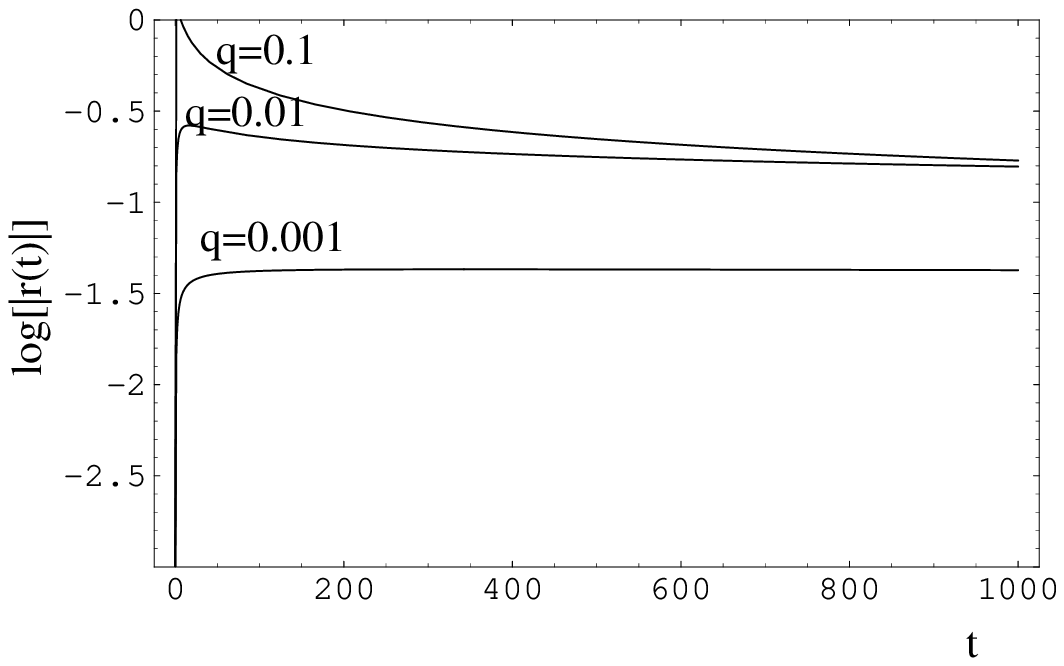}} \\
      \hline
\end{tabular}
\end{center}
\caption[a]{In this two plots we report the results of the integration of Eqs. 
(\ref{AI})--(\ref{AIII}) in the general case where magnetic field, dilaton and 
radiation fluid are present simultaneously. In these particular plot we assume 
that, initially, the dilaton  energy density is smaller (by two  orders 
of magnitude) than the radiation energy density. The left plot corresponds to 
the case of $\phi(t_0) =-0.2$ and the right plot  corresponds
 to the case of $\phi(t_0) = -2$. Different initial $q$ are reported in both 
cases. As in the previous cases we assumed $r(t_0)\sim 10^{-3}$.}
\label{f10}
\end{figure}

With the knowledge coming from the two previous examples we can easily 
analyze the general case given in Eqs. (\ref{AI})--(\ref{AIII}). The dynamics
 of the system will depend upon the different balance in the initial
 conditions. Let us suppose, firstly that the dilaton kinetic energy is 
slightly smaller (by two orders of magnitude) than the radiation energy 
density.
The results of the numerical integration of the shear parameter are reported
 in Fig. \ref{f10}. 
We can clearly see that by increasing the magnetic field the anisotropy 
will increase. The dilaton evolution clearly affects this process as we can 
argue by comparing Fig. \ref{f10} with Fig. \ref{f8} where the 
dialton was absent. It is also interesting to notice that the initial 
value of the coupling can be of some relevance. In Fig. \ref{f10} we reported
 the result of the integration for two values of the initial coupling, namely
$\phi_0=-0.1$ (at the left) and $\phi_0 = -2$ (at the right). The effect can 
be explained from Eqs. (\ref{AI})--(\ref{AIII}) where we can see that by tuning
the initial coupling to smaller values we amplify the effect of the magnetic 
energy density which appears always as $q e^{-\phi}$. 

\begin{figure}
\begin{center}
\begin{tabular}{|c|c|}
      \hline
      \hbox{\epsfxsize = 6.5 cm  \epsffile{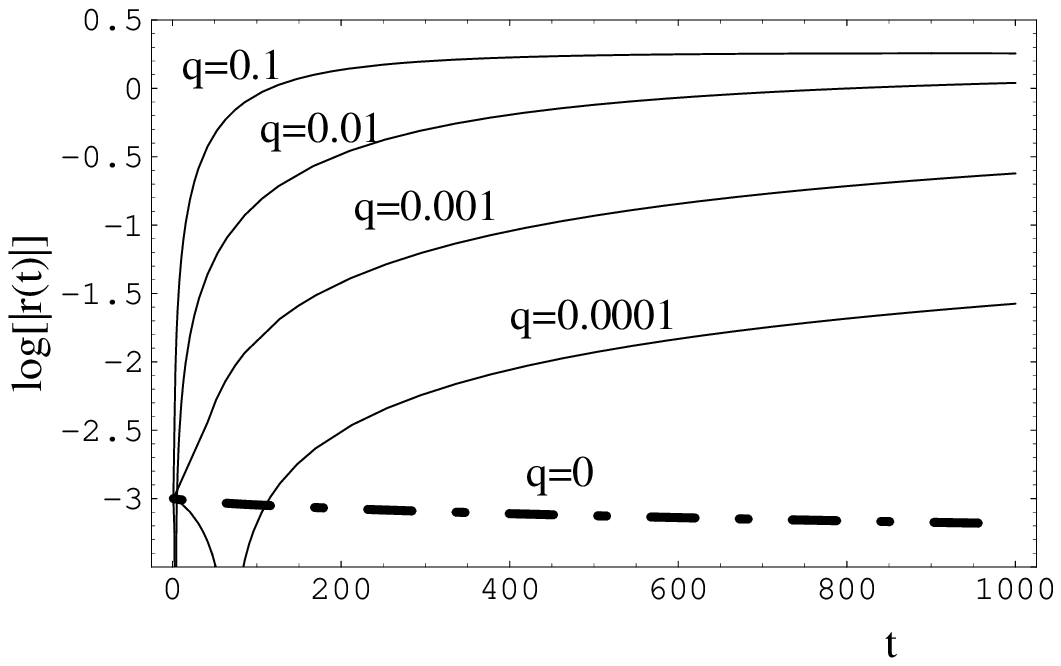}} &
      \hbox{\epsfxsize = 6.5 cm  \epsffile{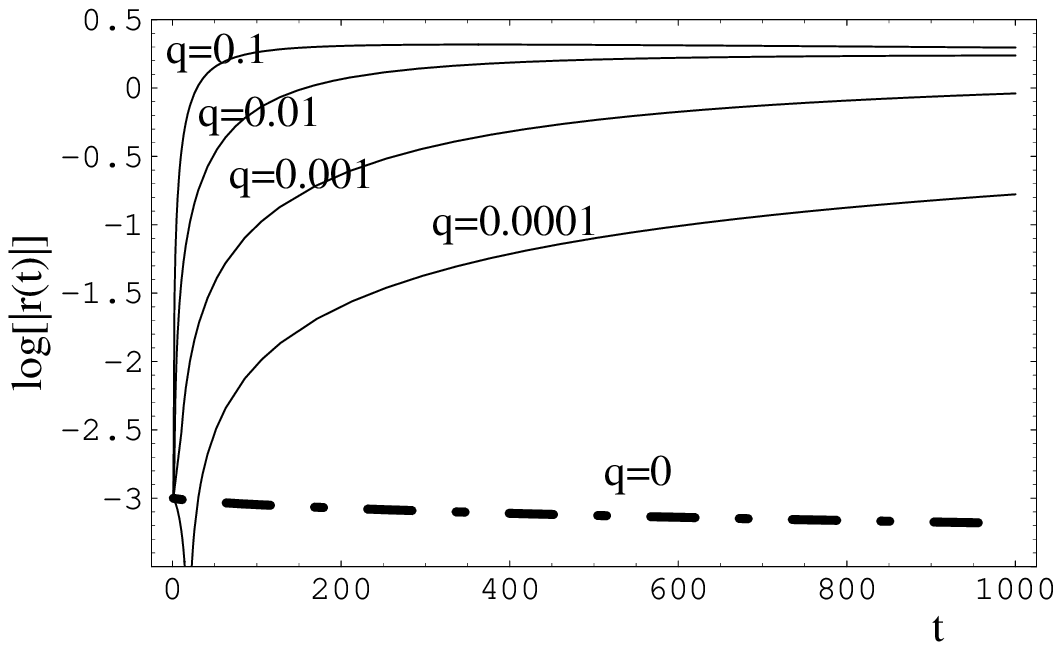}} \\
      \hline
\end{tabular}
\end{center}
\caption[a]{We report the numerical integration of the system given in 
Eqs. (\ref{AI})--(\ref{AIII}) for another set of parameters. In this case we
assume that the radiation energy density is hundred times smaller than the 
dilaton energy density and we watch the relaxation of the anisotropy 
for different values of the initial magnetic fields. As in the previous case 
the left picture refers to the case where $\phi(t_0) \sim -0.2$ whereas the 
right plot refers to the case where $\phi(t_0) \sim -2$. These plots have 
to be compared with the ones reported in Fig. \ref{f8} and Fig. \ref{f9}. 
We see
that, unlike in the case of Fig. \ref{f8}, the integration with $q(t_0)=0$ 
does not 
lead to an exactly constant anisotropy. For larger values of $q(t_0)$ the 
shear parameter saturates, as expected.}
\label{f11}
\end{figure}

If the radiation energy density is sub-leading with respect to the dilaton 
kinetic energy  we can expect, on the basis of the intuition developed 
in the previous cases that the shear  parameter will stay basically 
constant if
the magnetic field is turned off and it will increase to a constant value.
This is more or less what happens. In Fig. \ref{f11} we report the 
results of 
a numerical integration of the shear parameter in the case where the 
radiation energy is  hundred times smaller than the dilaton kinetic 
energy.
The left picture refers to the case of $\phi(t_0) \sim -0.2$ and the right 
figure refers to the case of $\phi(t_0) \sim -2$. In both cases 
we can notice
 that in the limit of zero magnetic field the anisotropy is not 
{\em exactly}
constant as in the case of $\rho=0$ reported in Fig. \ref{f9} but it has 
some mild slope and it decreases. At the same time, in contrast with 
Fig. \ref{f8} the asymptotic value of the anisotropy is not exactly 
the one we could guess on the basis of the argument illustrated in 
Fig. \ref{f8}.

\begin{figure}
\begin{center}
\begin{tabular}{|c|c|}
      \hline
      \hbox{\epsfxsize = 6.5 cm  \epsffile{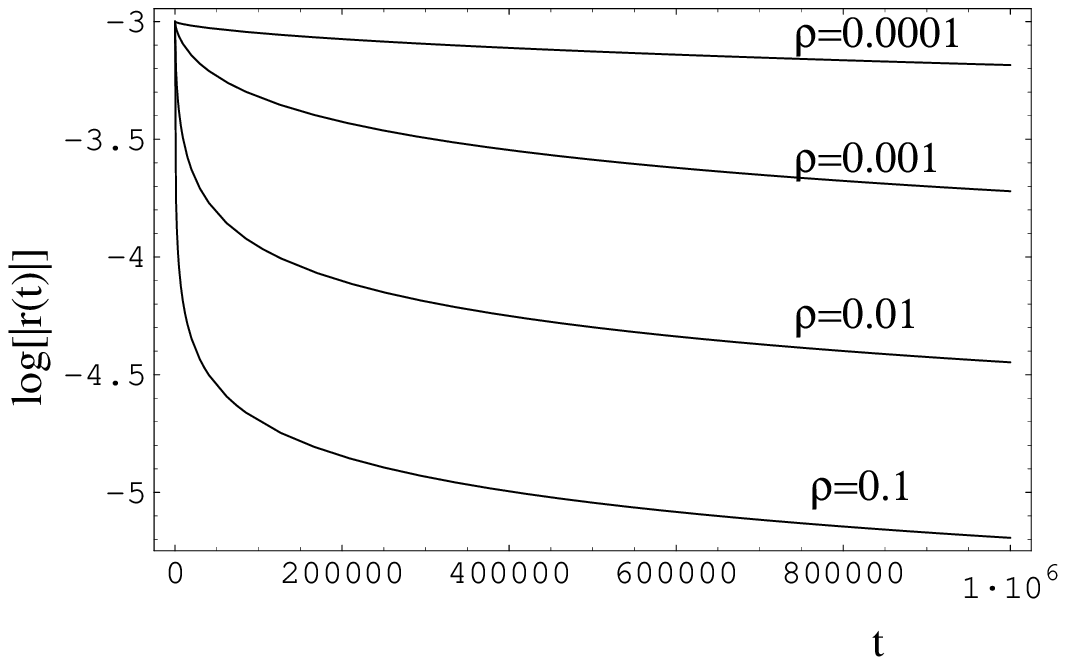}} &
      \hbox{\epsfxsize = 6.5 cm  \epsffile{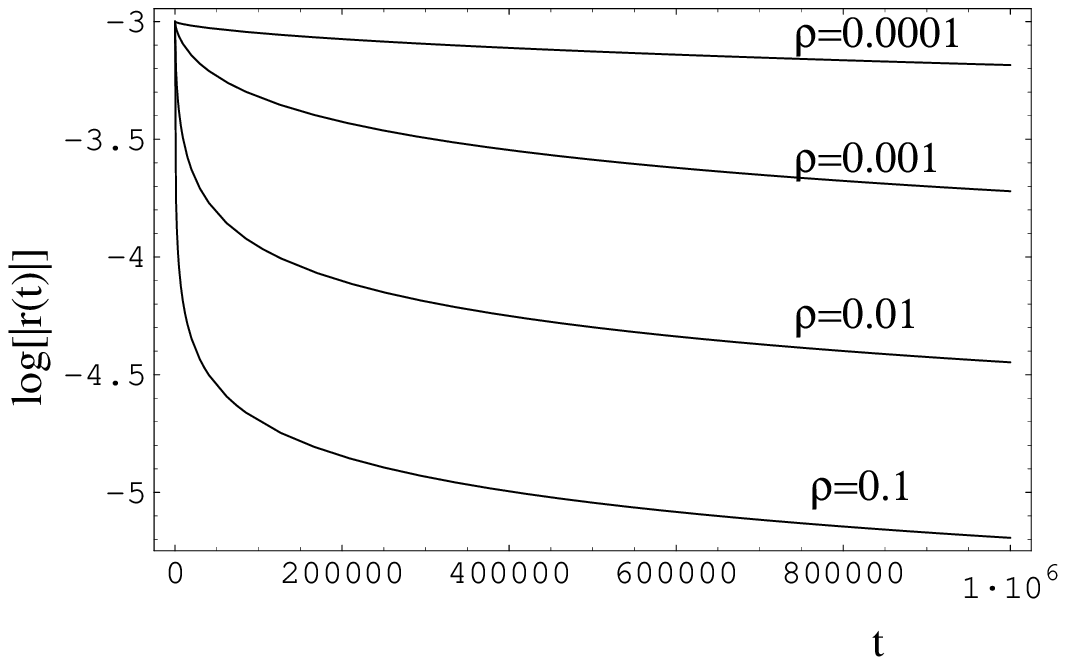}} \\
      \hline
\end{tabular}
\end{center}
\caption[a]{In this plots we assume that the magnetic field is initially 
zero. We 
also assume that the radiation energy density is smaller than the dilaton
energy density which is assumed to be of order one. The plot at the left 
is for 
$\phi(t_0) =-2$ whereas the plot at the right is for the case 
$\phi(t_0)=-10$. Nothing changes in the two plots  since, for 
both cases $q=0$.  
We notice that the behavior 
of the anisotropy is qualitatively similar both to the behaviors reported in 
Fig. \ref{f8} and Fig. \ref{f7} but with crucial differences. For instance,
 take the 
case with $\rho\sim 0.1$ at the left. In this case the radiation and dilaton 
energy densities are almost comparable. So there is some functional 
similarity with the plot of Fig. \ref{f7}. Look anyway at the time scale 
and take
 into account that the initial value of the anisotropy is the same 
in both cases. 
The same reduction of the shear parameter obtained in $100$ time steps 
in the case
 of pure radiation is now obtained in $10^{6}$ time steps. So the 
dilaton slows 
down the decay of the anisotropy in the absence of magnetic field. If the 
radiation energy density is gradually removed from the system the anisotropy 
is ( almost) constant.}
\label{f12}
\end{figure}

We want to stress, finally an important point. Suppose that
 the dilaton kinetic energy and the 
radiation energy are both present but the magnetic field is zero 
($q(t_0)=0$). Then we said that the anisotropy has a mild slope. 
The question is how mild. This point is addressed in Fig. \ref{f12} where 
the results of a numerical integration are reported for fixed dilaton 
kinetic energy and for radiation energy ten, hundred, thousand ad ten thousand
smaller than the dilaton kinetic energy.
We see that, in the best case, the anisotropy decrease of two orders of 
magnitude in $10^6$ time steps. This behavior should be compared with the 
sudden decrease experienced by the anisotropy in the case of Fig. \ref{f8} 
(dot-dashed line) where the dilaton was absent. Therefore, the presence 
of the dilaton in a radiation background slows down considerably, in the 
absence of magnetic field, the sudden fall-off of the shear parameter.

The conclusion we can draw from the analysis of the general case is that, also
in the presence of the dilaton field, the magnetic field has the effect of
increasing the anisotropy. Moreover, if the dilaton is larger than 
(or comparable with) the radiation fluid at the moment of the end of the 
string phase the anisotropy does decrease not so easily.

\renewcommand{\theequation}{6.\arabic{equation}}
\setcounter{equation}{0}
\section{Concluding Remarks} 

In this paper generalized the solutions of the low energy beta 
functions to the case where a magnetic field is present. We found that
these solutions can be expressed analytically in the String frame. 
We investigated their limit in the vicinity of the singularity and we 
noticed that they approach, monotonically, the well known 
Kasner-like vacuum solutions. 

We addressed a very simple question which can be 
phrased in the following way. Suppose that the dilaton-driven evolution
of the string cosmological models is fully anisotropic thanks, for 
instance, to a primordial magnetic seed. The obvious issue is 
to understand how (and possibly when) these fully anisotropic solutions
reach the complete isotropic stage. By complete isotropic stage we meant,
more quantitatively, a value of the shear parameter 
smaller, say, than $10^{-7}$. We reached a number of conclusions. 

For physical reasons we want to deal, from the very beginning, with 
anisotropic solutions whose scale factors are both expanding and accelerated. 
In this situation  it is not 
forbidden to have 
large anisotropies and shear parameters of order one 
prior to the onset of the string phase. 

Now, if the {\em string phase is sufficiently long} and if it is 
{\em immediately 
followed by a radiation dominated phase}, then, any pre-existing shear 
coming from the initial condition is very efficiently washed out to arbitrary 
small values depending upon the duration of the string phase. Of course, this 
conclusion holds {\em provided} no magnetic field is present in 
the ordinary decelerated phase. 
If a magnetic field is present, then the amount of 
shear 
appearing in our present Universe will not be determined by the primordial 
shear but by the shear {\em induced by the magnetic seed itself}. 
The role of the magnetic field might also be of some help but only in the case 
where the primordial shear is really large {\em after} the string phase. 
In this 
case a very small magnetic field can effectively attract the large 
anisotropy towards a smaller value. 

We can also have the opposite situation. Namely we can have the case where the 
{\em string phase is very short } and it is {\em followed by a phase dominated 
by the dilaton } until sufficiently small scales lasting from the end  of 
the sting phase until the onset of radiation which should anyway occur 
{\em before nucleosynthesis}. In this second scenario the shear 
will not be erased for two reasons. First of all because the duration 
of the string phase is proportional to the shear reduction. Secondly, 
because in a dilaton dominated phase (in the absence of magnetic field) 
the primordial shear is conserved.

\newpage

\begin{appendix}
\renewcommand{\theequation}{A.\arabic{equation}}
\setcounter{equation}{0}
\section{From the String to the Einstein frame}

In this Appendix we show explicitly how to get the Einstein 
frame action for our system. 
The transformation from the String to the Einstein frame {\em does  not}
remove the coupling of the dilaton to the kinetic term of the gauge fields.
The String frame action is simply
\begin{equation}
S= -  \int d^4 x \sqrt{- g} e^{-\phi} \biggl[ R + 
g^{\alpha\beta} \partial_{\alpha}\phi \partial_{\beta} \phi  
+ \frac{1}{4} F_{\alpha\beta}F^{\alpha\beta} \biggl].
\label{actionS}
\end{equation}
In four  dimensions the transformation from the String frame metric 
($g_{\mu\nu}$) to the Einstein frame metric ($G_{\mu\nu}$) simply reads 
\begin{equation}
g_{\mu\nu} = e^{\phi} G_{\mu\nu},~~~\sqrt{- g} = e^{ 2 \phi} \sqrt{- G}.
\end{equation}
Therefore,
\begin{eqnarray}
&&R= e^{-  \phi} \biggl[ {\cal R} 
- 3 G^{\alpha\beta} \nabla_{\alpha}\nabla_{\beta}\phi - \frac{3}{2} 
G^{\alpha\beta}\nabla_{\alpha}\phi\nabla_{\beta}\phi  \biggr],
\label{cov}\\
&&- \sqrt{- g} e^{-\phi} F_{\alpha\beta}F_{\rho\sigma} g^{\alpha\rho} 
g^{\sigma\beta} = - \sqrt{-G} e^{-\phi}F_{\alpha\beta}F_{\rho\sigma} 
G^{\alpha\rho} G^{\sigma\beta},
\end{eqnarray}
 the  derivatives in Eq. (\ref{cov}) 
are covariant with respect to the metric 
$G_{\mu\nu}$. Thus, the Einstein frame action can be written as 
\begin{equation}
S_{E} = \int d^4 x \sqrt{- G} \biggl[ - {\cal R} + \frac{1}{2} 
G^{\alpha\beta} \partial_{\alpha}\phi\partial_{\beta} \phi 
- \frac{1}{4} e^{- \phi} F_{\alpha\beta} F^{\alpha\beta}\biggr].
\label{actionE}
\end{equation}
Again from this equation we can derive the Equations of motion. We will just 
report the essential points. 
The Equations of motion derived from the action of Eq. (\ref{actionE}) read 
\begin{eqnarray}
&&{\cal R}_{\mu}^{\nu} 
- \frac{1}{2} \delta_{\mu}^{\nu} {\cal R}= \frac{1}{2} \biggl[T_{\mu}^{\nu}(d) 
+ e^{- \phi} T_{\mu}^{\nu}(m) + T_{\mu}^{\nu}(r)\biggr],
\nonumber\\
&& G^{\alpha\beta} \nabla_{\alpha} \nabla_{\beta}\phi = \frac{1}{4}
 e^{-\phi} F_{\alpha\beta} F^{\alpha\beta},~~~~ 
\nabla_{\alpha}\biggl[e^{-\phi} F^{\alpha\beta}\biggr] = 0
\end{eqnarray}
with
\begin{equation}
T_{\mu}^{\nu}(d) = \partial_{\mu} \phi \partial^{\nu}\phi + \frac{1}{2} 
\delta_{\mu}^{\nu} G^{\alpha\beta} \partial_{\alpha} \phi\partial_{\beta} 
\phi,~~~~ T_{\mu}^{\nu}(m) = - F_{\mu\alpha}F^{\nu\alpha} + \frac{1}{4} 
\delta_{\mu}^{\nu} F_{\alpha\beta}F^{\alpha\beta},~~~
 T_{\mu}^{\nu}(f) = {\rm diag}(\rho, -p, -p, -p),
\end{equation}
where on top of the energy momentum tensors of the dilaton 
(i.e. $T_{\mu}^{\nu}(d)$) and of the Maxwell fields (i.e.  
$T_{\mu}^{\nu}(m)$) we also added the energy momentum tensor of the 
fluid sources (i.e. $T_{\mu}^{\nu}(f)$)  
because of the considerations reported in Section V.
In the fully anisotropic metric given in Eq. (\ref{metric}) the previous 
equations of motion become
\begin{eqnarray}
&& F^2 + 2 H F = \frac{{\dot{\phi}}^2}{4} 
+  \frac{B^2}{4 b^4} e^{- \phi} + \frac{\rho}{2}, 
\label{E00}\\
&& 3 F^2 + 2 \dot{F} = - \frac{1}{4} {\dot{\phi}}^2 + 
\frac{B^2}{4 b^4}e^{- \phi} - \frac{p}{3},
\label{Exx}\\
&& \dot{H} + \dot{F} + H^2 + F^2 + HF = - \frac{{\dot{\phi}}^2}{4}
 - \frac{p}{2} - \frac{B^2}{4 b^4} e^{- \phi},  
\label{Eyy}\\
&& \ddot{\phi} + ( H + 2 F) \dot{\phi} = \frac{e^{-\phi}}{2} \frac{B^2}{b^4},
\label{Ephi}\\
&& \dot{\rho} + ( H + 2  F) ( \rho + p) =0.
\label{Erho}
\end{eqnarray}

Concerning these equations two technical comments are in order. 
First of all 
the over-dot denotes the derivation with respect to the {\em cosmic time of the 
Einstein frame} which is related to the cosmic time of the String frame 
(used, for instance in Section II) as 
\begin{equation}
d t_{E} = e^{- \phi/2} dt_{s},~~~
a_{E}(t_{E}) = e^{-\frac{\phi}{2}} a_{s}(t_s),~~~
b_{E}(t_{E}) = e^{-\frac{\phi}{2}} b_{s}(t_s).
\label{streintime}
\end{equation}
In the same way $H_{E}= (\log{a})^{\cdot}$ and 
$F_{E}=(\log{b})^{\cdot}$ are the Hubble
factors defined in the Einstein frame. We will not denote explicitly this 
distinction but we would like to remind 
 that the various quantities defined in Sections II, II, and IV are defined in 
the String frame whereas the discussion reported in Section V refers to the 
Einstein frame.

Most of the times one can check that all the ``physical'' quantities 
are invariant with respect to a change of frame. Typical examples 
of this property are the spectra of the fields excited by the 
dilaton growth in the case of the Kasner-like 
``vacuum'' solutions derived in Section II. Suppose in fact to compute 
the axionic (or gravitonic or dilatonic)  spectra in the String frame. Then 
Suppose to redo the same exercise in the Einstein frame with the transformed 
solutions. The two spectra will be equal. It is curious to notice 
that  the shear parameter is only approximately invariant under conformal 
rescaling. In order to show this aspect let us consider the dilaton-driven 
(vacuum) solutions reported in Eq. (\ref{solt}) and written in the String 
frame 
\begin{equation}
 a_{s}(t_{s}) = \biggl[-\frac{t_{s}}{t_1}\biggr]^{\alpha},~~~ b(t_{s}) = 
\biggl[-\frac{t_{s}}{t_1}\biggr]^{\beta},~~~\phi(t_{s}) = ( \alpha + 2 \beta
-1)\log{\biggl[-\frac{t_{s}}{t_{1}}\biggr]},
\label{vacstr}
\end{equation}
with $\alpha^2 + 2 \beta^2 =1$.
The shear parameter for this solution can be very simply computed in 
the string frame 
\begin{equation}
r(t_s) = \frac{ 3 [ H_{s}(t_s) - F_{s}(t_s)]}{ [H_{s}(t_s) + 2 F_{s}(t_s)]} 
\equiv  \frac{3(\alpha - \beta)}{ \alpha + 2 \beta}.
\label{rs}
\end{equation}
Let us therefore do the same exercise in the Einstein frame. 
The scale factors  and the cosmic time, transformed from the String to the 
Einstein frame are 
\begin{equation}
a_{E}(t_{E}) = \biggl[-\frac{t_{E}}{t_1}\biggr]^{\frac{\alpha - 2 \beta + 1}
{3 - \alpha - 2\beta} }, ~~~b_{E}(t_{E}) = 
\biggl[-\frac{t_{E}}{t_1}\biggr]^{\frac{1 - \alpha}{3 - \alpha - 2\beta} }
,~~~\biggl[ - \frac{t_s}{t_{1}}\biggr] \simeq 
\biggl[ - \frac{t_E}{t_{1}}\biggr]^{\frac{2}{3 - \alpha - 2\beta}}.
\end{equation}
Therefore we get that 
\begin{equation}
r(t_E) = \frac{ 3 [ H_{E}(t_E) - F_{E}(t_E)]}{ [H_{E}(t_E) + 2 F_{E}(t_E)]} 
\equiv  \frac{6(\alpha - \beta)}{ 3 -(\alpha + 2 \beta)}.
\label{re}
\end{equation}
Now, by comparing Eqs. (\ref{rs}) and (\ref{re}) it is clear that $r_s$ and 
$r_{E}$ go to zero in the same way and they are both proportional to
 $\alpha -\beta$. In this sense we can say that $r$ is {\em approximately}
invariant under conformal rescaling. For example by taking the anisotropic 
solution leading to flat axionic spectrum \cite{6}, namely $\alpha = - 7/9$, 
$\beta = - 4/9$ we get that $|r_{s}(t_s)|\sim 3/5$ whereas 
$|r_{E}(t_E)|\sim 3/7$. In all the cases discussed in this paper we 
explicitly checked that the shear parameters in one frame and in the other are 
of the same order of magnitude. So for practical purposes the shear parameters
computed in the either in the String or in the Einstein frame have the same 
quantitative information {\em provided} the calculation is performed with 
respect to  the correct cosmic time, which changes from frame to frame.
More generally one would like to find a frame-independent measure
 of the degree of anisotropy. One could argue that these quantity is provided 
by the Weyl tensor and, in particular by $C_{\mu\nu\alpha}^{~~~~\beta}$. Indeed
it is easy to show that $ C_{\mu\nu\alpha}^{~~~~\beta} = 
{\cal C}_{\mu\nu\alpha}^{~~~~\beta}$ where the calligraphic style denotes, as 
usual in this Appendix, the quantities computed in the Einstein frame. Notice 
that the index position is crucial \cite{wald} since 
$C_{\mu\nu\alpha\beta} = e^{\phi} {\cal C}_{\mu\nu\alpha\beta}$. The only 
problem with $C_{\mu\nu\alpha}^{~~~~\beta} $ is that it is dimension-full.
If we want to construct some dimension-less combination we run into the same 
ambiguity we just described since 
$C_{\mu\nu\alpha\beta}C^{\mu\nu\alpha\beta}/R_{\mu\nu\alpha\beta}
R^{mu\nu\alpha\beta}$ transforms non trivially under conformal rescaling. 
So our conclusion is that the shear parameter is perhaps still the best
 quantity in order to characterize the degree of anisotropy. In our problem,
moreover, we can show that the way the shear parameter goes to zero is a truly
frame independent statement since, ultimately if $r_{s}(t_s) 
\propto |\alpha - \beta|$ we also have that $r_{E}(t_E) 
\propto |\alpha - \beta|$.

The second point we would like to stress
concerns the covariant conservation of the (total) energy-momentum tensor in
the Einstein frame. The Bianchi identities impose 
\begin{equation}
\nabla_{\nu}\biggl[T_{\mu}^{\nu}(d) + e^{-\phi} 
T_{\mu}^{\nu}(m) + T_{\mu}^{\nu}(f)\biggr]=0.
\end{equation}
Now, it is easy to see that  the $0$ component of this equation,
in the case of a magnetic field directed along the $x$ axis, can be written as
\begin{equation}
(\partial_0 T_{0}^{0}(m) + 4 F T_{0}^{0}(m)) e^{- \phi} + 
\dot{\phi} \ddot{\phi} + {\dot{\phi}}^2 ( H + 2 F) - T_{0}^{0}(m) 
e^{-\phi}\dot{\phi}  + \dot{\rho} + ( H + 2 F) ( \rho + p) =0.
\end{equation}
Using now Eq. (\ref{Ephi}) we see immediately that the mixed term 
$T_{0}^{0}(m) e^{-\phi} \dot{\phi}$ cancels. Thus, in order to satisfy the 
covariant conservation of the energy-momentum tensor we have to impose  
\begin{equation}
\partial_{0}T_{0}^{0}(m) + 4 F T_{0}^{0}(m)=0,~~~~\dot{\rho} + (H+ 2 F)
 (\rho+  p) =0,
\end{equation}
which implies that $T_{0}^{0}= B^2/(2 b^4)$ (where $B$ is a constant) 
as correctly reported 
in Eqs. (\ref{E00})--(\ref{Erho}) 
from the very beginning as a consequence of the 
solutions of the equation for the field strength.

\end{appendix}

\end{document}